\documentclass[12pt,preprint]{aastex}

\usepackage{epsfig}

\def\Term#1 #2 #3/{\mbox{$\,^{#1}\!#2_{#3}$ }}
\def\Termo#1 #2 #3/{\mbox{$\,^{#1}\!#2^o_{#3}$ }}
\def\sterm #1 #2 #3/{\mbox{$\,_{#3}\!^{#1}\!#2$}}

\newcommand{\teff}{$T_{\rm eff}$} 

\begin{document}

\title{Spitzer Observations of M83 and the Hot Star, \ion{H}{2} Region Connection}

\author{Robert H. Rubin\altaffilmark{1,2},
Janet P. Simpson\altaffilmark{1,3},
Sean W.J. Colgan\altaffilmark{1},
Reginald J. Dufour\altaffilmark{4}, 
Katherine L. Ray\altaffilmark{1},
Edwin F. Erickson\altaffilmark{1}, 
Michael R. Haas\altaffilmark{1}, 
Adalbert W.A. Pauldrach\altaffilmark{5},
and
Robert I. Citron\altaffilmark{1}} 

\email{rubin@cygnus.arc.nasa.gov; simpson@cygnus.arc.nasa.gov}

\altaffiltext{1} {NASA/Ames Research Center, Moffett Field, CA
94035-1000, USA}
\altaffiltext{2} {Orion Enterprises, M.S. 245-6, Moffett Field, CA
94035-1000, USA}
\altaffiltext{3} {SETI Institute, 515 N. Whisman Road, Mountain View, CA
94043, USA}
\altaffiltext{4}{Physics \& Astronomy Department, Rice University, MS 61, 
Houston, TX 77005-1892, USA}
\altaffiltext{5}{University of Munich,  Munich D-81679, Germany}

\def\Termo#1 #2 #3/{\mbox{$\,^{#1}\!#2^o_{#3}$ }}

\date{\today}

\begin{abstract}
We have undertaken a program to observe emission lines of 
[\ion{S}{4}] 10.51,
[\ion{Ne}{2}] 12.81,
[\ion{Ne}{3}] 15.56,
and [\ion{S}{3}] 18.71~$\mu$m
in  a number of
extragalactic \ion{H}{2} regions with the Spitzer Space Telescope. 
Here we report our results for the nearly face-on spiral galaxy M83.
A subsequent paper will present our data and analysis for another substantially
face-on spiral galaxy M33.
The nebulae selected cover a wide range of galactocentric radii (R$_G$).  
The observations were made with the Infrared Spectrograph in the 
short wavelength, high dispersion configuration.
The above set of four lines is observed cospatially, 
thus permitting a reliable comparison of the fluxes.
 From the measured fluxes,
we determine the ionic abundance ratios including
Ne$^{++}$/Ne$^+$,
S$^{3+}$/S$^{++}$, and S$^{++}$/Ne$^+$ 
and find that there is a correlation of increasingly higher ionization with 
larger R$_G$.
By sampling the dominant ionization states of Ne and S
for \ion{H}{2} regions,  we can approximate
the Ne/S ratio
by (Ne$^+$ + Ne$^{++}$)/(S$^{++}$ + S$^{3+}$).
Our findings of ratios that significantly exceed the benchmark
Orion Nebula value, as well as a decrease in this ratio with increasing R$_G$,  
are more likely due to other effects than a true gradient in Ne/S.
Two effects that will tend to lower these high estimates and
to flatten the gradient are first,  the method 
does not account for the presence of S$^+$ and
second, S but not Ne is incorporated into grains.
Both Ne and S are primary elements produced in $\alpha$-chain reactions, 
following C and O burning in stars,
making their yields depend very little on the stellar metallicity. 
Thus, it is expected that Ne/S remains relatively constant throughout a galaxy.
We stress that this type of observation and method of analysis
does have the potential for accurate measurements of Ne/S,
particularly for \ion{H}{2} regions that have lower metallicity and 
higher ionization than those here,
such as those in M33.   
	Our observations may also be used to test 
the predicted ionizing spectral energy distribution (SED) of various 
stellar atmosphere models.
We compare the ratio of fractional ionizations
$<$Ne$^{++}$$>$/$<$S$^{++}$$>$ 
and 
$<$Ne$^{++}$$>$/$<$S$^{3+}$$>$
vs.\ 
$<$S$^{3+}$$>$/$<$S$^{++}$$>$
with predictions
made from our photoionization models using 
several of the state-of-the-art stellar atmosphere model grids.
The overall best fit appears to be the nebular models using the
supergiant stellar atmosphere models of Pauldrach~et~al.\ (2001) 
and  Sternberg, Hoffmann, \& Pauldrach (2003).
This result is not sensitive to the electron density and temperature range
expected for these M83 nebulae.
Considerable computational effort has gone into the comparison between 
data and models, although not all parameter studies have yet been
performed on an ultimate level (e.g., in the present paper the 
stellar atmosphere model abundances have been fixed to solar values). 
A future paper, with the benefit of more observational data,
will continue these studies to further discriminate how the
ionic ratios depend on the SED and the other nebular parameters.
\end{abstract}


\vspace{12pt}

\section{Introduction}

     Most observational studies of the chemical evolution of the universe
rest on emission line objects, which define the mix of elemental abundances at
advanced stages of evolution as well as 
the current state of the interstellar medium (ISM).
Gaseous nebulae 
are laboratories for understanding physical 
processes in all emission-line sources and probes for stellar, 
galactic, and primordial nucleosynthesis. 
\ion{H}{2} regions are also among the best tracers of recent star formation.

The presence of radial (metal/H) abundance gradients in the plane of the Milky Way 
is well established in both gaseous nebulae and stars
(e.g., Henry \& Worthey 1999; Rolleston~et~al.\ 2000).
Radial abundance gradients 
seem to be ubiquitous 
in spiral galaxies, 
though the degree varies
depending on a given spiral's morphology and luminosity class. 
The gradients are 
generally attributed to the radial dependence of star formation 
history and ISM mixing processes
(e.g., Shields 2002).
Thus, the observed gradients are a major tool for understanding galactic 
evolution 
(e.g., Hou~et~al.\ 2000; Chiappini et~al.\ 2001; Chiappini et~al.\ 2003).
The premise 
is that star formation and chemical enrichment begins
in the 
nuclear bulges of the galaxies and subsequently 
progresses outward into 
the disk, which has remained gas-rich.
The higher molecular gas density in the inner regions produces a higher star 
formation rate, which results in a 
relatively greater return to the ISM
of both ``primary'' 
$\alpha$-elements (including O, Ne, and S)
from massive star supernovae, 
and ``secondary'' elements like N. 
Secondary nitrogen is produced by CNO burning of already existing carbon and oxygen 
in intermediate-mass stars and 
is subsequently returned to the ISM through mass loss.
However, because chemical evolution models have
uncertain input parameters, and because
details of the abundance variations of each element
are uncertain,
current understanding of the formation and evolution of galaxies suffers
(e.g., Pagel 2001).

	Studies of \ion{H}{2} regions
in the Milky Way are hampered by interstellar extinction.
For the most part, optical studies (e.g., Shaver~et~al.\  1983) 
have been limited to those \ion{H}{2} regions
at galactocentric radius R$_G \gtrsim 6$ kpc
(predicated on $R_\odot = 8$ kpc) 
because \ion{H}{2} regions are very concentrated to the Galactic plane.
Here extinction becomes severe with increasing distance from Earth.
Observations using far-infrared (FIR) emission lines
have penetrated the R$_G \lesssim 6$~kpc barrier. 
Surveys with the Kuiper Airborne Observatory (KAO)
by Simpson~et~al.\ (1995),
Afflerbach~et~al.\ (1997),
and Rudolph~et~al.\ (2006)
have observed 16 inner Galaxy \ion{H}{2} regions.
With the Infrared Space Observatory (ISO),
Mart\'{\i}n-Hern\'{a}ndez~et~al.\ (2002a) observed 13 inner Galaxy 
\ion{H}{2} regions covering FIR and also mid-IR lines.
A major finding of these studies is that inner Galaxy \ion{H}{2} 
regions generally have lower excitation (ionization)
compared to those at larger R$_G$.
This holds for both heavy element ionic ratios O$^{++}$/S$^{++}$ 
(Simpson~et~al.\ 1995) and 
Ne$^{++}$/Ne$^{+}$ (Simpson \& Rubin 1990; Giveon~et~al.\ 2002),
and also He$^+$/H$^+$ measured from radio recombination lines 
(Churchwell~et~al.\ 1978; Thum, Mezger, \& Pankonin 1980).
Whether the observed increase in  excitation with increasing R$_G$ 
comes  entirely from heavy element opacity effects 
in the \ion{H}{2} regions and stellar atmospheres,
or also from 
a gradient in the maximum stellar effective temperature, \teff, 
of the exciting stars 
is still a point of controversy
(e.g., Giveon~et~al.\ 2002; 
Mart\'{\i}n-Hern\'{a}ndez~et~al.\ 2002b; 
Smith, Norris, \& Crowther 2002; Morisset~et~al.\ 2004).

	It has become clear that 
nebular plasma simulations with photoionization modeling codes
are enormously sensitive
to the ionizing spectral energy distribution (SED) that is input
(e.g., Simpson et~al.\ 2004, and references therein).
These SEDs need to come from stellar atmosphere models.
Stellar atmosphere modelers 
are paying increasing attention 
to the usefulness of nebular observations,
particularly of \ion{H}{2} regions, in validating and constraining their
models (e.g.,  Stasi\'nska \& Schaerer 1997).
Though it may be somewhat chauvinistic to say so, 
much of this increased attention
stems from work we did pointing out the 
\underbar{``[\ion{Ne}{3}] problem"} and possible solutions
(Simpson~et~al.\ 1995; Rubin~et~al.\ 1995; Sellmaier~et~al.\ 1996).
To produce Ne$^{++}$ requires ionizing photons $> 41$ eV.
At energies exceeding this ionization potential (IP), 
modern stellar atmosphere model SEDs are especially divergent.
A critical test of the validity of stellar atmosphere models of hot stars is
whether \ion{H}{2} region models produced with these atmospheres 
predict line fluxes that agree with observations.
The ``[\ion{Ne}{3}] problem" is that the observed Ne$^{++}$/O$^{++}$ ratio 
significantly exceeds model predictions and remains
relatively constant over a large range of \ion{H}{2} region excitation as 
gauged by the O$^{++}$/S$^{++}$ ratio.
The original observational basis is KAO FIR measurements 
of [\ion{Ne}{3}] 36,
[\ion{O}{3}] 52, and 
[\ion{S}{3}] 33~\micron ~lines in Galactic \ion{H}{2} regions
(see figure~3 in Simpson~et~al.\ 1995).

While Sellmaier~et~al.\ (1996) believed that 
they had solved the [\ion{Ne}{3}] problem 
when they obtained a good fit to the FIR data by
using non-LTE atmospheres with winds computed with Pauldrach's code as it then 
existed, we recently investigated 
the problem and  concluded that it still exists (Simpson~et~al.\ 2004).
Our \ion{H}{2} region models using non-LTE stellar atmospheres with 
winds from 
Pauldrach, Hoffmann, \& Lennon (2001)
and from atmosphere models by both Smith~et~al.\ (2002) 
and Sternberg, Hoffmann, \& Pauldrach (2003)
using Pauldrach's WM-BASIC code predict much lower Ne$^{++}$/O$^{++}$ 
than observed
for ``Dwarf" atmospheres with \teff\ $< 40,000$ K and for 
``Supergiant" atmospheres 
with \teff\ $< 35,000$ K. 
Moreover, the models without winds 
do a poorer job of reproducing the observations than those that include winds.

In this  paper, we develop new observational tests of and constraints 
on the ionizing SEDs that are predicted from various stellar atmosphere models.
We do this by utilizing Spitzer Space Telescope (SST) observations of
\ion{H}{2} regions in the spiral galaxy M83.
M83 (NGC~5236) is one of the closest (distance 3.7~Mpc)
and brightest spirals (SBc~II),
as well as being nearly face-on ($i$~= 24$^{\rm o}$)
(de~Vaucouleurs et~al.\ 1983).
M83 particularly interested us because of its high metallicity 
(at least twice solar, e.g., Dufour et~al.\ 1980;
Bresolin \& Kennicutt 2002). 
High metallicity should correlate with lower-ionization \ion{H}{2} regions.
   From our experience with the [\ion{Ne}{3}] problem,
this is likely to provide the most stringent test matching 
theory with observations.
With the Infrared Spectrograph (IRS) on the SST,
we can observe four emission lines that
probe the dominant ionization states of neon and sulfur
in these \ion{H}{2} regions.
These lines are:
[\ion{Ne}{2}] 12.81,
[\ion{Ne}{3}] 15.56,
[\ion{S}{3}] 18.71, and [\ion{S}{4}] 10.51~\micron.
Crucially, SST permits the {\bf simultaneous, cospatial}
observation of these four lines.

	We discuss the SST/IRS observations in section 2. 
In section 3, the data are used to test for
a variation in the degree of ionization of the
\ion{H}{2} regions with R$_G$.
We examine the Ne/S abundance ratio for our 
M83 \ion{H}{2} region sample in section 4.
Section 5 describes how these Spitzer data are used to  constrain
and test the ionizing SEDs predicted by stellar atmosphere models.
Last, we provide a summary and conclusions in section~6.

\section{Spitzer Space Telescope Observations}

	In the nearly face-on (tilt 24$^{\rm o}$)
spiral galaxy M83,
we observed 24 \ion{H}{2} regions, 
covering a wide range of deprojected galactocentric radii (R$_G$). 
We used the SST/IRS in 
the short wavelength, high dispersion (spectral resolution $\sim$ 600) 
configuration,
called the short-high (SH) mode
(e.g.,  Houck~et~al.\ 2004).
This covers the wavelength range from 9.9~-- 19.6~$\mu$m
permitting cospatial observations of all four of our program emission lines:
[\ion{S}{4}] 10.51,
[\ion{Ne}{2}] 12.81,
[\ion{Ne}{3}] 15.56,
and [\ion{S}{3}] 18.71~\micron.
The SH bandpass also covers the hydrogen 
Hu$\alpha$, also called  H(7--6),  at  12.37~\micron.
Unfortunately, we were not able  to detect this line in any
of the 24 \ion{H}{2} regions observed.

	The observations were made in 2005, February and July.
Figure~1 shows the regions and apertures observed,
while Table~1 lists the \ion{H}{2} region positions
and the aperture grid configuration used to observe it.
Nebulae with RK and deV designations are from Rumstay \& Kaufman (1983) 
and  de~Vaucouleurs~et~al.\ (1983), respectively.
The size  of the SH aperture is 11.3$''$$\times$4.7$''$. 
In all cases, we chose the mapping mode  with 
aperture grid patterns varying from a
1$\times$2 grid to as  large as a 2$\times$4 grid
in order to 
cover the bulk of the emission.
Maps were arranged with the apertures immediately
abutting each other; that is, with no overlap or space between them.
In order to save overhead time, we clustered the objects into 
``Astronomical Observing Templates" (AOTs) with the same 
aperture grid pattern.

	Our data were processed and calibrated 
with both versions S12.0.2  and S13.2.0 of
the standard IRS pipeline at the Spitzer Science Center (SSC).
For each position, we have 16 cycles with 30 sec ramp duration.
The basic calibrated data (bcd images) for each telescope pointing
were median-combined and cleaned of rogue pixels and noisy
order edges (the ends of the slit).
The spectra were extracted and the lines measured using the IRS
Spectroscopic Modeling, Analysis and Reduction Tool 
(SMART, Higdon~et~al.\ 2004).
For the brighter lines, we find little change in the
line fluxes between the two  pipeline versions.
Generally the [\ion{S}{4}] 10.5 line was weakest;
for these we used the later version of the pipeline.
The emission lines were measured with SMART using a Gaussian line fit.
The continuum baseline was fit with a linear or
quadratic function.
Figures~2 (a)--(d) show the fits for each of the four lines
in RK209 (object \#9 in Fig.~1).
A line is deemed to be detected if the flux is  at least
as  large as the 3~$\sigma$ uncertainty.
We measure the uncertainty by the product of the FWHM  and
the  root-mean-square
variations in the adjacent, line-free continuum;
it does not include systematic  effects.

We now discuss and estimate systematic 
uncertainties for our specific case using IRS in SH mode.
Most likely the largest uncertainty is due to slit (aperture) loss factors.
The pipeline flux calibration assumes that objects are point sources.
Our nebulae are extended and that is why  we mapped each with a grid  that
covers more than  a single aperture.  
We did not make a correction for this effect.
Thus we have implicitly assumed that the \ion{H}{2} regions are 
close to the point-source limit within the SH 11.3$''$$\times$4.7$''$ aperture.
If the \ion{H}{2} region were uniformly extended within the SH aperture,
correction factors would need  to  be applied  to our  fluxes.
These are: 0.697, 0.663, 0.601, and 0.543 for the 
10.5, 12.8, 15.6, and 18.7~$\mu$m lines respectively
(Simpson et~al.\  in preparation).
These factors were obtained by interpolating in numbers provided from 
the $b1\_slitloss\_convert.tbl$ file from the
Spitzer IRS Custom Extraction tool (SPICE) for the SH module (see
http://ssc.spitzer.caltech.edu/postbcd/spice.html).
For the uniformly filled aperture,  the 
maximum uncertainty in the flux due to this effect
would be $\sim$46\% for the [\ion{S}{3}] 18.7 line.  
The correction factors would need to multiply our listed fluxes.
We note that with regard to this effect, the fluxes listed in Table~2
are  upper limits and that the uncertainty would be in only the
direction to lower them.
No correction factor was applied because we are likely closer to
the point-source limit than the uniform-brightness limit.
Because our science depends on line flux ratios, for our purposes, the 
possible  uncertainty due to this effect would be lower, e.g.,
$\sim$22\% when we deal with the line flux ratio 
[\ion{S}{4}] 10.5/[\ion{S}{3}] 18.7.

	According to \S7.2 {\it  Spectroscopic Flux Calibration
Uncertainties} (Infrared Spectrograph Data Handbook, ver.\ 2.0
(http://ssc.spitzer.caltech.edu/irs/dhb),
the minimum uncertainty possible in the absolute
flux calibration of the spectroscopic products delivered by the pipeline
is $\pm$5\%  due to the ``photometric uncertainty introduced by 
uncertainty in the angular sizes of the standard stars and their spectral 
types".  It is also stated that the maximum uncertainty [for a point source]
is $\pm$10\%  by comparisons with other Spitzer instruments.

	Any uncertainty in the flux due to a pointing error is probably small
and in the worst case should not exceed 10\%.  We arrive at this estimate
as follows.
The absolute pointing accuracy of Spitzer is $\sim$1$''$ rms. 
Since a spectral map is performed by alignment of a guide star, this is 
effectively the pointing uncertainty for each spatial position in the spectral map.
Because our smallest map dimensions are 11.3$''$$\times$ 9.4$''$,
an error of 1$''$ could result in an error of $\sim$10\% if the source
uniformly filled  the  map  area  and went to zero outside of  it.

For the brighter lines, i.e., most of the 12.8, 15.6, and 18.7~$\mu$m lines,
the systematic uncertainty far exceeds the measured (statistical) uncertainty.
Even for the fainter lines, we estimate that the
systematic uncertainty exceeds the measured uncertainty.

	In addition to the line flux, the measured FWHM and
radial velocity (V$_r$) are listed in Table~2.
Both the FWHM and V$_r$ are useful in judging the
reliability of the line measurements. 
The FWHM is expected to be the instrumental width
for all our lines.  With a resolving power for the
SH module of $\sim$600, our lines should have a
FWHM of roughly 500~km~s$^{-1}$.
The values for V$_r$ should straddle the heliocentric systemic radial
velocity for M83 of $\sim$516~km~s$^{-1}$
(Lawrence~et~al.\  1999). 
We note that the IRS pipeline does not correct 
to heliocentric radial velocities (V$_{helio}$).
At the low ecliptic latitude of M83, the 
correction of V$_r$ to V$_{helio}$ can be almost
$\pm$30~km~s$^{-1}$.
It is interesting (but not pertinent)
that we can definitely measure
an average velocity differential between our February
and July data of roughly 60~km~s$^{-1}$, as expected.

\section{Variation in the degree of ionization of the
\ion{H}{2} regions with R$_G$}

     From the measured fluxes, we estimate ionic 
abundance ratios, including  Ne$^{++}$/Ne$^+$,
S$^{3+}$/S$^{++}$, and S$^{++}$/Ne$^+$, 
for each of the \ion{H}{2} regions.
Important advantages compared with prior optical studies 
of various other ionic ratios are: 
1) the IR lines have a weak and similar electron temperature ($T_e$) 
dependence while the collisionally-excited
optical lines vary exponentially with $T_e$, and 
2) the IR lines suffer far less from interstellar extinction.  
Indeed for our purposes,  the differential extinction correction
is negligible as the lines are relatively close in wavelength.
In our analysis, we deal with ionic abundance ratios
and therefore line flux ratios.
In order to derive the 
ionic abundance ratios, we perform the usual semiempirical
analysis assuming a constant $T_e$ and electron density ($N_e$)
to obtain the volume emissivities for the four pertinent transitions.
We use the  atomic data described
in Simpson et~al.\ (2004)
and 
Simpson et~al.\ (in preparation)
  for the ions Ne$^+$,
Ne$^{++}$, S$^{++}$, and S$^{3+}$. 
For the entries in Table~3, we 
adopt a typical value for all the M83 \ion{H}{2} regions
of $T_e$~= 8000~K and $N_e$~= 100~cm$^{-3}$.
The \ion{H}{2} regions in M83 are  known to have a high metallicity 
(e.g., at least twice solar, Dufour et~al.\ 1980;
Bresolin \& Kennicutt 2002); 
thus a value of 8000~K is not unreasonable.
This is also typical of the values found for seven
\ion{H}{2} regions in M83 (see Tables~10 and 11 in Bresolin et~al.\ 2005).
Because  of the insensitivity of the volume emissivities
to $T_e$, particularly when working with ratios for these IR lines,
our results depend very little on this $T_e$ choice. 
The effects on our analysis due to a change in the assumed
$N_e$ are also small as will be discussed later.

	We chose our sample of nebulae 
in order to cover a wide range in R$_G$.  
It is a straight-forward geometry exercise to derive the deprojected
galactocentric distances.
This involves knowing the inclination angle ($i$~= 24$^{\rm o}$),
position angle of the line of nodes  ($\theta = 43^{\rm o}$),
and distance (D~= 3.7~Mpc) (de~Vaucouleurs~et~al.\ 1983).
We assumed the centre of the galaxy is at
$\alpha$, $\delta$ = $13^{\rm h}37^{\rm m}00\fs92$, 
$-29^{\rm o}$51\arcmin56\farcs7 (J2000) 
(source NASA/IPAC Extragalactic Database).
Table~3 lists R$_G$ for the centre of each object.
These range from 0.46 to 5.16~kpc.

	We present the variation of Ne$^{++}$/Ne$^+$ with R$_G$
in Figure~3 using the values from Table~3.
The error values represent the propagated flux measurement uncertainties and 
do not include the systematic uncertainties.
There is extremely little change in any of these ratios
even when using an $N_e$  of 1000~cm$^{-3}$, which is likely 
a reasonable upper limit for these \ion{H}{2} regions
(see Figure~9 in Bresolin et~al.\ 2005).
A linear least-squares fit indicates a positive
correlation with R$_G$ (in kpc),

~~~~~~~~~~~~~~~Ne$^{++}$/Ne$^+$~= 0.033$\pm$0.010~+~(0.011$\pm$0.0035)~R$_G$,

\noindent
with miniscule change to this equation for $N_e$~= 1000~cm$^{-3}$.
For all the least-squares line fits in this paper,
each point is given equal weight
because systematic uncertainties exceed the flux measurement 
uncertainties, as discussed earlier.
The positive correlation of Ne$^{++}$/Ne$^+$  
with R$_G$ as measured by the slope may
be judged to be significant following the criterion that it
exceeds the 3~$\sigma$ uncertainty.

	A similar fit to the 	
S$^{3+}$/S$^{++}$  vs.\ R$_G$ data yields

~~~~~~~~~~~~~~~S$^{3+}$/S$^{++}$~= $-$0.00079$\pm$0.0079~+~(0.0067$\pm$0.0027)~R$_G$.

\noindent
There is  more  scatter in the plot (not shown)
because the data for the [\ion{S}{4}] 10.5 line are noisier than for the
other stronger lines (see Figure~2).
The slope exceeds  the 1~$\sigma$ uncertainty
by a factor of only 2.4.
Thus the increase in degree of ionization with increasing
R$_G$ in this case would be deemed marginal.
We note that one of the points (for RK268) stands out
as far above any of the others.
If we refit eliminating this far outlier, then

~~~~~~~~~~~~~~~S$^{3+}$/S$^{++}$~= 0.0055$\pm$0.0032~+~(0.0029$\pm$0.0011)~R$_G$.

\noindent
The slope is not as steep and as 2.5~$\sigma$ would again 
be deemed marginally significant.

	Figure~4 plots
the fractional ionic abundance ratio
$<$S$^{++}$$>$/$<$Ne$^+$$>$ vs.\ R$_G$. 
This ratio is obtained from the 
S$^{++}$/Ne$^+$ ratio by multiplying by an assumed Ne/S value (see below).
The last three columns of Table~3 list this and other
fractional ionic abundance ratios used in this paper.
In Figure~4,
the filled circles
represent the points, and the solid line,
the linear least-squares fit for an assumed $N_e$ of 100~cm$^{-3}$.
For \ion{H}{2} regions in M83 this is likely the typical
$N_e$ value.
Here,  the fit indicates a significant positive correlation with R$_G$,

~~~~~~~~~~~~~~~$<$S$^{++}$$>$/$<$Ne$^+$$>$~= 0.43$\pm$0.026~+~(0.035$\pm$0.0089)~R$_G$, 

\noindent
where angular brackets denote fractional ionization.
The lower dashed line is
the least-squares fit 
to points marked  with an X derived 
assuming $N_e$~= 1000~cm$^{-3}$.
This shows the effect of higher $N_e$
on the volume emissivity of the [\ion{S}{3}] line.
The least-squares fit for this density is

~~~~~~~~~~~~~~~$<$S$^{++}$$>$/$<$Ne$^+$$>$~= 0.36$\pm$0.022~+~(0.030$\pm$0.0076)~R$_G$,

\noindent
and here too, the slope is statistically significant.
In this figure and in these linear fits, we assume an Orion Nebula Ne/S 
abundance ratio of 14.3 (Simpson et~al.\ 2004).
Because Ne and S are ``primary" elements,
their production is expected to vary in lockstep
and Ne/S would not be expected to show a radial gradient within a galaxy
(Pagel \& Edmunds 1981).
There is a clear correlation of increasingly higher ionization with 
increasing R$_G$.
This is most likely due to the lower metallicity at larger R$_G$
causing the exciting stars to have a harder ionizing spectrum.  
The first quantitative abundance study of radial gradients
for \ion{H}{2} regions  in M83
found dlog(O/H)/dR(kpc)~= $-0.09\pm0.02$~dex~kpc$^{-1}$
based on differential
photoionization modeling (Dufour et~al.\ 1980). 
With $\sim$25 more years of  observations, this gradient
still appears but has flattened considerably.
In their Figure 7, Bresolin \& Kennicutt (2002)
plot the O/H gradient.
When we convert their units to R(kpc) using 
the M83 distance of 3.7 Mpc,
we find 
dlog(O/H)/dR(kpc)~= $-$0.0257~dex~kpc$^{-1}$.
A shallower slope is  also indicated in Figure 15 
(see also their Figure 20) in Bresolin  et~al.\ (2005),
where the M83 \ion{H}{2} region  points are shown as open squares.
In their  Figure 15, M83 has the flattest O/H gradient among the
5 galaxies plotted.
Because  M83 has a bar, this tends to correlate with
a flatter radial abundance gradient (e.g., Martin \& Roy 1994)
due to radial mixing.

\section{Neon to Sulfur abundance ratio}

For \ion{H}{2} regions, we may approximate the Ne/S ratio with
(Ne$^+$ + Ne$^{++}$)/(S$^{++}$ + S$^{3+}$).
This includes the dominant ionization states of these two elements.
However this relation does not account for S$^+$,
which should  be present at some level.
We may safely ignore the negligible contributions
of neutral Ne and S in the ionized region.
Figure~5 shows our approximation for Ne/S vs.\ R$_G$.
There appears to be a drop in the Ne/S ratio
with increasing R$_G$.
The expected increasing fraction of
S$^+$ towards the inner galaxy regions would
lead to a flatter gradient.
Another factor that could flatten the slope is the higher dust content
(with S, but  not Ne, entering grains)
expected in the inner regions due to higher metallicity
as is the case for the Milky Way.
The refractory carbonaceous and silicate grains are not distributed
uniformly throughout the Galaxy but instead increase in density toward the
centre.
A simple  model suggests the dust density is $\sim$5~-- 35 times higher
in the inner parts  of the Galaxy than in the local ISM 
(Sandford,  Pendleton,  \& Allamandola 1995).
The Ne/S abundance ratios that we derive here are considerably 
higher than  the Orion Nebula value
of 14.3 (Simpson~et~al.\ 2004). 
We suspect that all our Ne/S estimates are upper limits
because of the two effects (not accounting for S$^+$ or dust)
with the outer regions likely needing less of a downward
correction to obtain a true Ne/S ratio.

	The premise that these estimates of Ne/S 
are upper limits 
is further supported by our forthcoming work with 
\ion{H}{2} regions in M33.
This is a similar study for which we 
have Spitzer  Cycle 2
spectra  of roughly 25 nebulae in
the substantially face-on  spiral  M33.
A preliminary report was presented at IAU Symposium 235
(Rubin et~al.\ 2006).
The  Ne/S values (under the same
approximation) fell mostly in the range from
$\sim$12~-- 21.
M33 \ion{H}{2} regions have  a lower metallicity than those in M83.
Also, most of the \ion{H}{2} regions we observed in M33 have 
significantly higher ionization than those we observed in M83.
These two facts tend to mitigate 
the amount of any downward correction
needed to account for S$^+$ and dust.
While there appears to be a decrease in 
our (approximate) Ne/S vs.\ R$_G$  for M33,
our data also  indicate that the  lower envelope to  Ne/S
is well fit by a constant value equal to  the Orion Nebula ratio
(Rubin et~al.\ 2006 and in preparation).

\section{Constraints on the ionizing SED for the stars exciting the
\ion{H}{2} regions}

	Various fractional ionic abundances are 
highly sensitivity to the stellar ionizing SED that apply to \ion{H}{2} regions.
This has been realized, for example, for the Ne$^+$~-- Ne$^{++}$
ionization equilibrium and the total number of photons more energetic
than the 41~eV Ne$^+$ ionization potential that are predicted by various
stellar atmosphere models (e.g., Simpson~et~al.\ 2004 and references therein).
The present Spitzer data probe the 
Ne$^+$ and Ne$^{++}$ fractional ionic abundances,
as well as those of  
S$^{++}$ and S$^{3+}$;
thus they may be used to provide further constraints  and tests 
on the ionizing SED for the stars exciting these M83 nebulae.
We use the 
ratio of fractional ionizations
$<$Ne$^{++}$$>$/$<$S$^{++}$$>$ 
vs.\ 
$<$S$^{3+}$$>$/$<$S$^{++}$$>$
(Figure~6a) 
and 
$<$Ne$^{++}$$>$/$<$S$^{3+}$$>$
vs.\ 
$<$S$^{3+}$$>$/$<$S$^{++}$$>$
(Figure~6b). 
These ionic ratios are computed 
using our photoionization code NEBULA
(e.g., Simpson et~al.\  2004; Rodr\'\i guez \& Rubin 2005).
The lines connect the results of the nebular models calculated using the 
ionizing SEDs predicted from various stellar atmosphere models.
There are \underbar{no other changes} to the input parameters,
just the SED.
The stellar atmospheres used are  representative of several
recent non-LTE models that apply for O-stars.
We also  display the results from one set of LTE models
by Kurucz (1992).  His LTE atmospheres have been extensively used in
the past as input for \ion{H}{2} region models;
hence the comparison with the other non-LTE results reinforces the fact 
that more reliable SEDs for O-stars require a non-LTE treatment.
Figures~6a and b dramatically illustrate how sensitive 
\ion{H}{2} region model predictions of these
ionic abundance ratios are to the
ionizing SED input to nebular plasma simulations.

	For the \ion{H}{2} region models calculated with 
Pauldrach~et~al.\ (2001) atmospheres, 
the solid line connects models with dwarf atmospheres and the dashed line 
connects models with supergiant atmospheres.
The Sternberg et~al.\ (2003) paper also uses Pauldrach's WM-BASIC code.
At a given  \teff\,
we have used their model with the smallest log~$g$ in order to
be closest to the supergiant case.
Because the locus using these Sternberg et~al.\ atmosphere models
is for the most part similar to the Pauldrach~et~al.\ supergiant locus, 
we do not show it in Figures~6  to avoid clutter.
For \ion{H}{2} region models calculated with Lanz \& Hubeny (2003)
atmospheres  (TLUSTY code),
the solid line connects models with atmospheres with log~$g= 4.0$
and the dotted and dashed lines connect models with atmospheres with
log~$g= 3.0$ to 3.5, and with Lyman continuum luminosities of $10^{49}$
and $10^{50}$ photons s$^{-1}$, respectively.
The lines with open squares in Figures~6 are the results of our
nebular models with the atmospheres in Martins~et~al.\ (2005) 
that use Hillier's CMFGEN code.

	To compare our data with the models, we need to divide the observed 
Ne$^{++}$/S$^{++}$
and
Ne$^{++}$/S$^{3+}$
ratios by an assumed Ne/S abundance ratio.
For the purposes of Figure~6,  we 
adopt a constant Ne/S~= 14.3, the Orion Nebula 
value  (Simpson~et~al.\ 2004).
As per the discussion in \S4, we cannot conclude 
definitively whether Ne/S may vary.
The open circles 
(adjusted by the assumed Ne/S)
are derived from our observed line fluxes
using $N_e$ of 100~cm$^{-3}$.
The M83 data usually
lie closest to the Pauldrach~et~al.\ 
supergiant loci.
In addition the points derived from our data as well as  those for
the \ion{H}{2} regions we observed  in M33 (Rubin~et~al.\ 2006 and in 
preparation), which are generally of higher ionization,
follow the trend of these theoretical loci.
This is particularly notable in Figure~6b, where the other model 
loci are nearly perpendicular to the data point trend in the 
vicinity of where they intersect the data points.

	The nebular models used to generate Figures~6
are all constant density, spherical models.
We used a constant total nucleon density (DENS)  of 1000~cm$^{-3}$
that begins at the star.
Each model used a total number of Lyman continuum photons~s$^{-1}$
($N_{Lyc}$)~= 10$^{49}$.
The same {\it nebular} elemental abundance set was used for all 
nebular models.
We use the same ``reference" set as in Simpson et al.\ (2004) because in
that paper  we were studying the effects of various  SEDs  on
other ionic ratios and other data sets.
Ten elements are included with their abundance  by number
relative to H as follows:
(He, C, N, O, Ne, Si, S, Ar, Fe) with
(0.100, 3.75E$-$4, 1.02E$-$4, 6.00E$-$4, 1.50E$-$4, 2.25E$-$5, 1.05E$-$5, 
3.75E$-$6, 4.05E$-$6), respectively.
While the purpose here was not to try to match the 
abundances in  the \ion{H}{2} regions we observed  in 
M83, the set of abundances used is roughly a factor of 
1.5 higher than for Orion and not drastically different from solar.
We have investigated the effects of changing DENS, $N_{Lyc}$, 
and allowing for a central  evacuated cavity, 
characterized by an  initial radius (R$_{init}$)
before the stellar radiation encounters nebular material.
We term these shell models.
In Figures~7a and 7b, the resulting changes to Figures~6a and 6b
are  shown for six nebular models run using two of
the Pauldrach ~et~al.\ (2001) supergiant atmospheres.
These are listed along with the symbol:

\noindent
(1) 
\teff\ = 35000 K; DENS = 1000~cm$^{-3}$; R$_{init}$ = 0.5 pc;  $N_{Lyc}$ = 
10$^{49}$~s$^{-1}$ (asterisk)

\noindent
(2)
\teff\ = 35000 K; DENS = 100~cm$^{-3}$; R$_{init}$ = 0 pc;  $N_{Lyc}$ = 
10$^{49}$~s$^{-1}$ (triangle)

\noindent
(3)
\teff\ = 35000 K; DENS = 100~cm$^{-3}$; R$_{init}$ = 0 pc;  $N_{Lyc}$ = 
10$^{50}$~s$^{-1}$ (X)

\noindent
(4)
\teff\ = 40000 K; DENS = 1000~cm$^{-3}$; R$_{init}$ = 0.5 pc;  $N_{Lyc}$ = 
10$^{49}$~s$^{-1}$ (asterisk)

\noindent
(5)
\teff\ = 40000 K; DENS = 100~cm$^{-3}$; R$_{init}$ = 0 pc;  $N_{Lyc}$ = 
10$^{49}$~s$^{-1}$ (triangle)

\noindent
(6)
\teff\ = 40000 K; DENS = 100~cm$^{-3}$; R$_{init}$ = 0 pc;  $N_{Lyc}$ = 
10$^{50}$~s$^{-1}$ (X)

	The original Pauldrach ~et~al.\ (2001) supergiant locus
and points derived from the Spitzer data are shown again.
We also display the effects of using a different $N_e$
to interpret our line measurements in terms of ionic ratios.
The open stars show the results 
with $N_e$ of 1000~cm$^{-3}$ while the original points
(open circles) were derived  with $N_e$~= 100~cm$^{-3}$.
The change is slight with the higher $N_e$ shifting points to the upper right
in Figure~7a and to the lower right in Figure~7b.

	The points  for cases (3)  and (6) are nearly identical to the 
original points for the Pauldrach supergiant models at the same 
respective \teff.
This can be understood in terms of the ionization
parameter ($U$), which is very useful for gauging ionization
structure.
An increase in $U$ corresponds to  higher ionization (for a given
$T_{\rm eff}$). 
For an ionization bounded, constant density case,
$$U = [N_e N_{Lyc} (\alpha - \alpha _1)^2/(36 \pi c^3)]^{1/3}~~,$$
\noindent
where ($\alpha$ - $\alpha$$_1$) is the recombination rate coefficient to 
excited levels of hydrogen, and $c$ is the velocity of light
(see Rubin~et~al.\ 1994,  eq.\ 1 and adjoining discussion).
Because\break
($\alpha$ - $\alpha$$_1$)~$\simeq$ 
4.10$\times$10$^{-10}~T_e^{-0.8}$ 
cm$^{3}$~s$^{-1}$ (Rubin 1968 fit to Seaton 1959),
there is only a weak dependence of $U$ on $T_e$ ($\sim$ $T_e^{-0.5}$).
When $U$ is similar, as is the case here with the product of 
$N_e\times$$N_{Lyc}$, the ionization structure is similar.
	With regard to the two shell models  in Figures~7,
the  Str\"omgren radius is $\sim$0.74~pc.
Thus the radial thickness of the shell is slightly less  than half
the radius of the central cavity.  From the visual appearance of our
target \ion{H}{2} regions,
it is unlikely that the theoretical loci need to be tracked to higher
dilutions.

	Another {\it nebular} parameter that can alter the 
theoretical tracks is the set of elemental abundances used.
This is certainly well established.
For instance, in the grid of model \ion{H}{2} regions of Rubin (1985),
the fractional ionic abundances needed here were tabulated
considering a variation in the heavy-element abundance set
of a factor of 10.
The up (``U")  and down (``D") sets in that paper
were meant to represent $\sqrt{10}$ and 1/$\sqrt{10}$
times the nebular (``N") Orion-like abundances at that time.
A sampling of those models, that are closest to the parameter space
of interest in this paper, indicates median shifts of the following factors:
$<$Ne$^{++}$$>$/$<$S$^{++}$$>$~= 3.50,
$<$Ne$^{++}$$>$/$<$S$^{3+}$$>$~= 1.39,
and
$<$S$^{3+}$$>$/$<$S$^{++}$$>$~= 2.13,
when going from ``U" to ``D" sets.
Much more apropos for the current situation, we have calculated two variants
of the Pauldrach \teff\ = 40000~K supergiant (canonical model in Figures 6 
and 7) with the only change being in the {\it nebular} abundance set by scaling 
all the heavy elements by a factor of two higher and two lower than
the set used for all other models.
This is more than sufficient to cover the expected variation in our
M83 sample allowing for the heavy-element gradient
dlog(O/H)/dR(kpc)~= $-$0.0257~dex~kpc$^{-1}$ (see \S3).
As expected, lower metallicity results in a shift to higher
ionization.  In both Figures 6a,b, the point moves to
the upper right;
the factors are 
$<$Ne$^{++}$$>$/$<$S$^{++}$$>$~= 1.29,
$<$Ne$^{++}$$>$/$<$S$^{3+}$$>$~= 1.16,
and
$<$S$^{3+}$$>$/$<$S$^{++}$$>$~= 1.11.
Likewise, higher metallicity shifts the point to the lower left.
This latter change is relatively larger with factors:
$<$Ne$^{++}$$>$/$<$S$^{++}$$>$~= 1.62,
$<$Ne$^{++}$$>$/$<$S$^{3+}$$>$~= 1.26,
and
$<$S$^{3+}$$>$/$<$S$^{++}$$>$~= 1.28.
In the case of the M83 nebulae, we may conclude
that the predicted spread in Figures 6 due to a  reasonable
uncertainty in nebular metallicity is  far less than that due
to the SEDs of the various stellar atmosphere models.

	There is also the effect of a change in the abundances used to 
compute the stellar atmosphere models.
This will change the emergent stellar SED (e.g., Mokiem et~al.\ 2004).
As is the case  for a change in  the nebular model abundance set,
such a modification in the stellar model will alter the shape 
of the SED in the same sense;
that is, a higher metallicity will cause more opacity and soften the
SED, and a lower metallicity will do the opposite.
Mokiem et~al.\ (2004) examined this using CMFGEN stellar models
matching both the nebular and stellar metallicities. 
Their Figure~11 tracks the predicted variation in the 
[\ion{Ne}{3}] 15.6/[\ion{Ne}{2}] 12.8 flux ratio
over a range of 0.1~-- 2~Z$_{\sun}$.

Although beyond the scope of this paper, it would be interesting to compare 
models using different stellar atmosphere metallicities, especially if the 
environment indicates significant departures from solar. 
However, presently the proper abundances are not
accurately known and comparisons like those presented in this paper 
will help decipher the proper values.
Improvements with regard to this point are deferred for a future paper, 
where the curves for different metallicities are compared with the observations. 
With the present paper, however, we have been investigating 
which of the models best fits the observations and 
such a comparison works most effectively with a fixed set 
of abundances {\it common to all models}, which are the solar ones.

Such additional degrees of freedom (besides the SED) 
in the nebular models
makes judging how well
a particular stellar atmosphere model set
fits observations more challenging.
We will revisit these fits in a second paper that
has the benefit of additional  observational data
of \ion{H}{2} regions in M33 (Rubin~et~al.\ in preparation).

\section{Summary and conclusions}

	We have observed emission lines of 
[\ion{Ne}{2}] 12.81,
[\ion{Ne}{3}] 15.56,
[\ion{S}{3}] 18.71, and [\ion{S}{4}] 10.51~$\mu$m
cospatially with the Spitzer Space Telescope using the Infrared
Spectrograph (IRS) in short-high mode (SH).  From the measured fluxes,
we determined the ionic abundance ratios Ne$^{++}$/Ne$^+$,
S$^{3+}$/S$^{++}$, and S$^{++}$/Ne$^+$ 
in 24 \ion{H}{2} regions in the substantially face-on spiral galaxy M83.  
These nebulae cover a range from 0.46 to 5.16~kpc
in deprojected galactocentric distance  R$_G$.
We found a correlation of increasingly higher ionization with increasing R$_G$.
This is seen in the variation of 
Ne$^{++}$/Ne$^+$ and $<$S$^{++}$$>$/$<$Ne$^+$$>$ with R$_G$ (see  Figures 3 and  4).  
This is most likely due to the lower metallicity at larger R$_G$
causing the exciting stars to have a harder ionizing spectrum.

By sampling the dominant ionization states of Ne and S
for \ion{H}{2} regions, we can approximate the Ne/S ratio
by (Ne$^+$ + Ne$^{++}$)/(S$^{++}$ + S$^{3+}$).
The decrease in this ratio with increasing R$_G$
is more likely due to other effects than a true
gradient in Ne/S.
Both Ne and S are the products of 
$\alpha$-chain reactions
following carbon and oxygen burning in stars,
with large production factors from core-collapse supernovae.
Both are primary elements, making their yields depend very
little on the stellar metallicity. 
Thus, at least to ``first order", it is expected that Ne/S remains 
relatively constant throughout a galaxy.

As discussed in \S4, our estimate for Ne/S
has not accounted for the presence of S$^+$.
Because inner-galaxy \ion{H}{2} regions have lower ionization
(\S3), it is possible that there would be a larger fractional ionization of
S$^+$ towards the inner galaxy regions and 
hence a flatter Ne/S slope than  indicated in Figure~5.
A second factor which might flatten this apparent gradient
is the higher grain content in the inner regions due to higher metallicity.
All of our
derived Ne/S abundance ratios are considerably 
higher than the benchmark Orion Nebula value of 14.3.
Because we did not account for S$^+$ or dust
(with the objects at larger R$_G$ likely needing less of a downward
correction to obtain a true Ne/S ratio),
our Ne/S estimates may be considered upper limits.
This conclusion is further supported by our subsequent
work on \ion{H}{2} regions in M33 where
this methodology leads to lower estimates for Ne/S.
This was briefly discussed in \S4 with
citations to a conference proceedings and a future
paper (Rubin~et~al.\ 2006 and in press).
Thus with observations of the set of four IR emission lines
and the analysis we have described, there is the potential 
for reliable Ne/S measurements, especially for \ion{H}{2} regions 
that have lower metallicity  and higher ionization than those in M83.

	At the present time, the solar abundance, particularly of Ne, 
is the  subject of much controversy (e.g., Drake \& Testa 2005; Bahcall, 
Serenelli, \& Basu 2006;
and references in each of these).
While we cannot directly address the solar abundance with our
observations of extragalactic \ion{H}{2} regions,
it is important to  have reliable benchmarks for the neon abundance.
There appears to be a growing  body of evidence that the Ne abundance
[its fractional number abundance relative to hydrogen
log~H~= 12, by definition and termed $A$(H)]
is substantially higher in the solar neighborhood, and even in the Sun
itself, than the ``canonical" solar values 
given  in two recent,  often-referenced papers.
These  papers   have for the Sun:
$A$(Ne)~= 7.87, $A$(S)~= 7.19  (Lodders 2003)
and
$A$(Ne)~= 7.84, $A$(S)~= 7.14  (Asplund, Grevesse,  \& Sauval 2005).
Thus  Ne/S~$\sim$ 5 
according to both.
It is now generally accepted that Ne has the least
well determined solar abundance among the most abundant elements.
One of the proponents for a higher Ne abundance
pointed out that an $A$(Ne)~= 8.29 would reconcile solar models with 
the helioseismological  measurements (Bahcall, Basu, \& Serenelli 2005).
Using this value  together with the $A$(S) values above,
we obtain Ne/S of 12.6 and 14.1,  respectively,
close  to  the Orion Nebula ratio (Simpson~et~al.\ 2004).

	According to calculations based on
the theoretical  nucleosynthesis, galactic
chemical  evolution models  of Timmes, Woosley,  \& Weaver (1995),
the Ne/S ratio in the solar neighborhood would change little,
from 3.80 to  3.75, between solar birth and the present
time  (apropos  for the Orion Nebula).
These calculations were provided by Frank Timmes (private communication).
With regard to these ratios being  even lower than the
canonical solar values, 
Timmes notes that although massive stars
are expected to dominate the Ne and S production (and are all that
were included in their non-rotating models with no wind losses),
there is likely to be some re-distribution of Ne and S from rotation 
or from Wolf-Rayet phases of evolution, along with contributions 
of Ne from novae or even heavier intermediate mass stars.
Hence, the theoretical model Ne/S ratios above should
be a lower bound because some potential sources of Ne are missing.

	Additionally, the  data here may be used as constraints on the ionizing SEDs  
for the stars exciting these nebulae by comparing the ratio of fractional ionizations
$<$Ne$^{++}$$>$/$<$S$^{++}$$>$ 
and 
$<$Ne$^{++}$$>$/$<$S$^{3+}$$>$
vs.\ 
$<$S$^{3+}$$>$/$<$S$^{++}$$>$
with predictions
made from our photoionization models using stellar atmosphere models
from several different sources.
Figures~6 show the comparison, where we 
assume that the Ne/S ratio does not vary and equals the Orion Nebula value.
Generally, the best fit is to the nebular models using the
supergiant stellar atmosphere models 
(Pauldrach~et~al.\ 2001)
computed with the WM-BASIC code.
We note that this comparison is mainly qualitative
since these ionic ratios depend not only on the SED, but also
on the nebular parameters discussed as well
as the effects of the stellar metallicity on the SED.
This result is not sensitive to the electron density range, 
100 and 1000~cm$^{-3}$, expected for these M83 nebulae.
Furthermore, the points derived from our M83 data, as well as  those for
the \ion{H}{2} regions we observed  in M33 (Rubin~et~al.\ 2006 and in 
preparation), which are generally of higher ionization,
follow the trend of the above-mentioned  theoretical loci.
In fact, 
the other model loci are nearly perpendicular  to 
the data point trend in the vicinity of where they intersect the data points
in Figure~6b.
We reiterate that we do not infer that the actual exciting stars 
are supergiants, but only that their SEDs have similar shape as 
the supergiant atmospheres  computed using WM-BASIC.

It is possible that a plot similar to Figures 6a,b using
$<$Ne$^{++}$$>$/$<$Ne$^+$$>$ for the ordinate will also have 
utility in comparing data with models using different SEDs.
While the $<$Ne$^{++}$$>$/$<$Ne$^+$$>$ ratio has the advantage of
being independent of elemental abundance ratios, it appears to be 
more sensitive to the {\it nebular} parameters than does the 
$<$Ne$^{++}$$>$/$<$S$^{3+}$$>$ ratio.
This fact tends to make it less unique in its ability to discriminate
between the stellar SEDs we present in this paper.
We defer this investigation until our next paper that has the addition 
of more Spitzer data with \ion{H}{2} regions in M33.

\begin{acknowledgments}
This work is based on observations made with the Spitzer Space Telescope,
which is operated by the Jet Propulsion Laboratory, California Institute
of Technology under NASA contract 1407.  Support for this work was provided
by NASA for this Spitzer program identification 3412.
RHR had further support from the NASA Long-Term Space Astrophysics (LTSA)
program.
We thank 
Frank Timmes for providing information 
on the Ne/S ratio from a nucleosynthesis,
galactic chemical evolution perspective.
We thank Brendan Wakefield and Danny Key for assistance with the data reduction.
Valuable comments by the referee are much appreciated.
The nebular models were run on a Cray computer at JPL.
Funding
for its use in this investigation was provided by 
the JPL Office of the Chief Information Officer.
\end{acknowledgments}

\begin{figure}
\centering
%        \resizebox{16.0cm}{!}{\includegraphics{m83.fig1.ps}}
\resizebox{16.0cm}{!}{\includegraphics{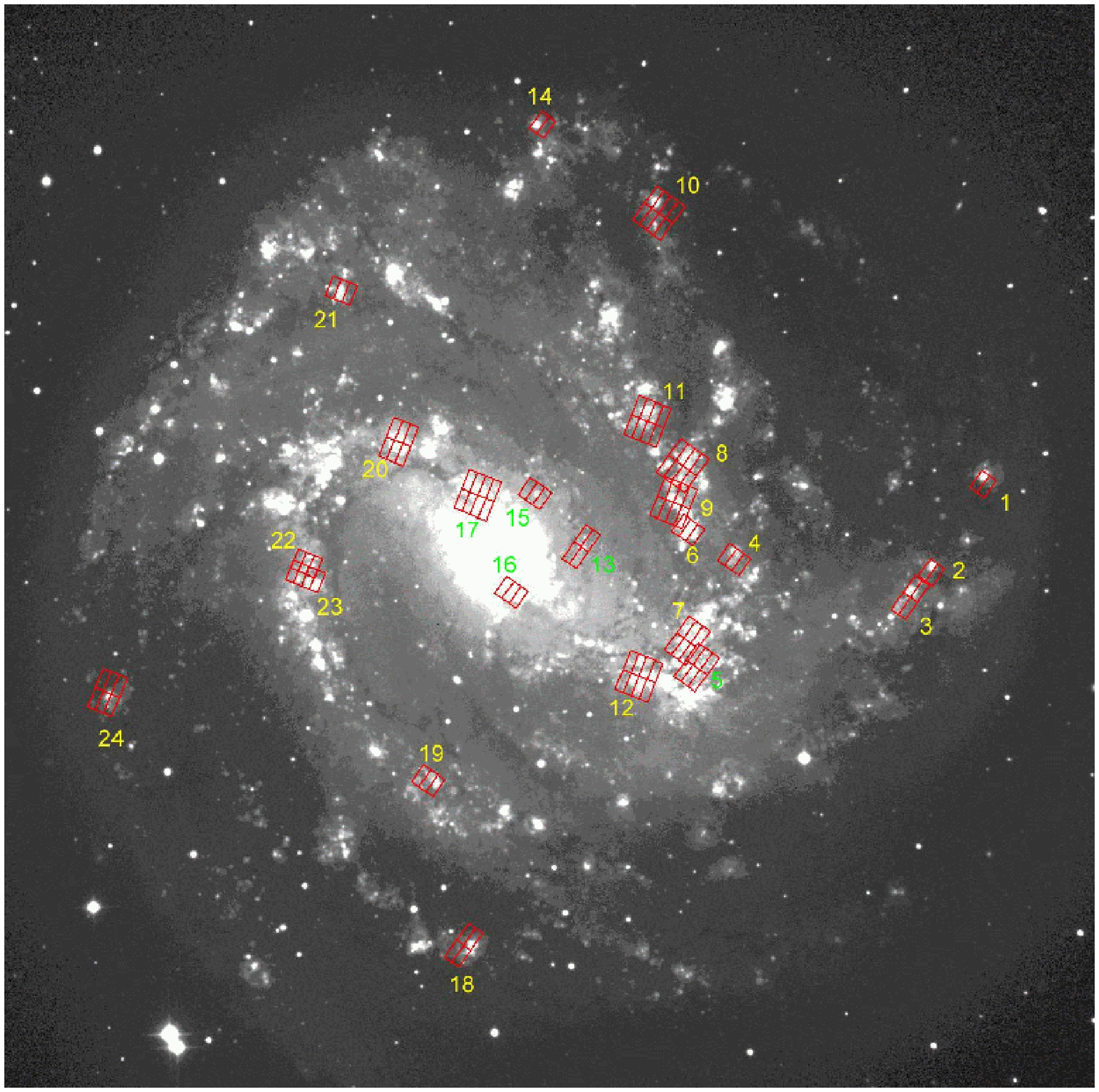}}
\vskip-0.1truein
\caption[]{The positions and apertures observed for 24 \ion{H}{2} regions
are shown in red superimposed on an H$\alpha$ image of 
the nearly face-on (tilt 24$^{\rm o}$)
M83.
The nebulae are numbered W to E (see Table~1).
N is up and E left.}
\end{figure}

\begin{figure}
  \begin{center}
    \begin{tabular}{cc}
      \resizebox{80mm}{!}{\includegraphics{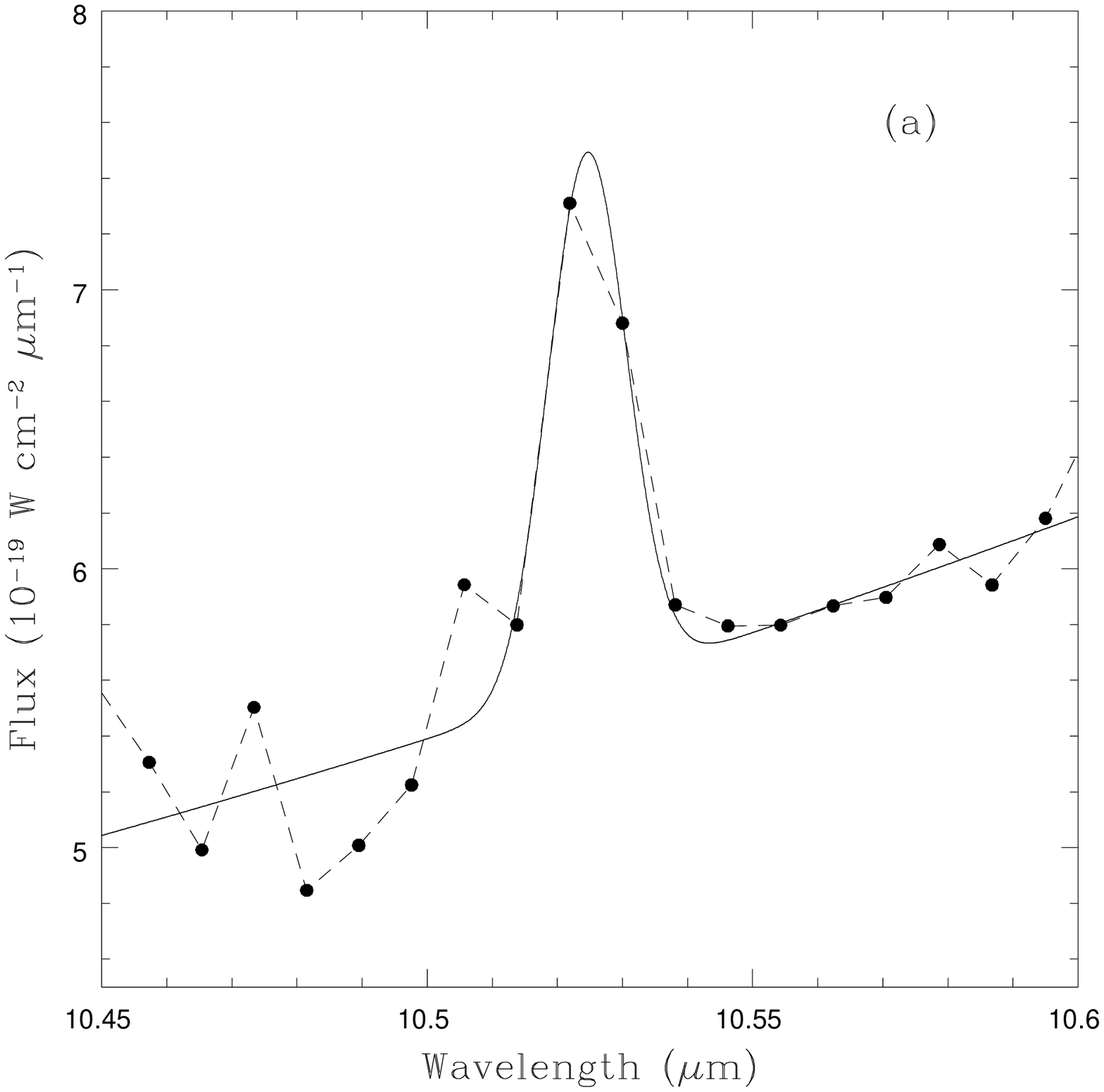}} &
      \resizebox{80mm}{!}{\includegraphics{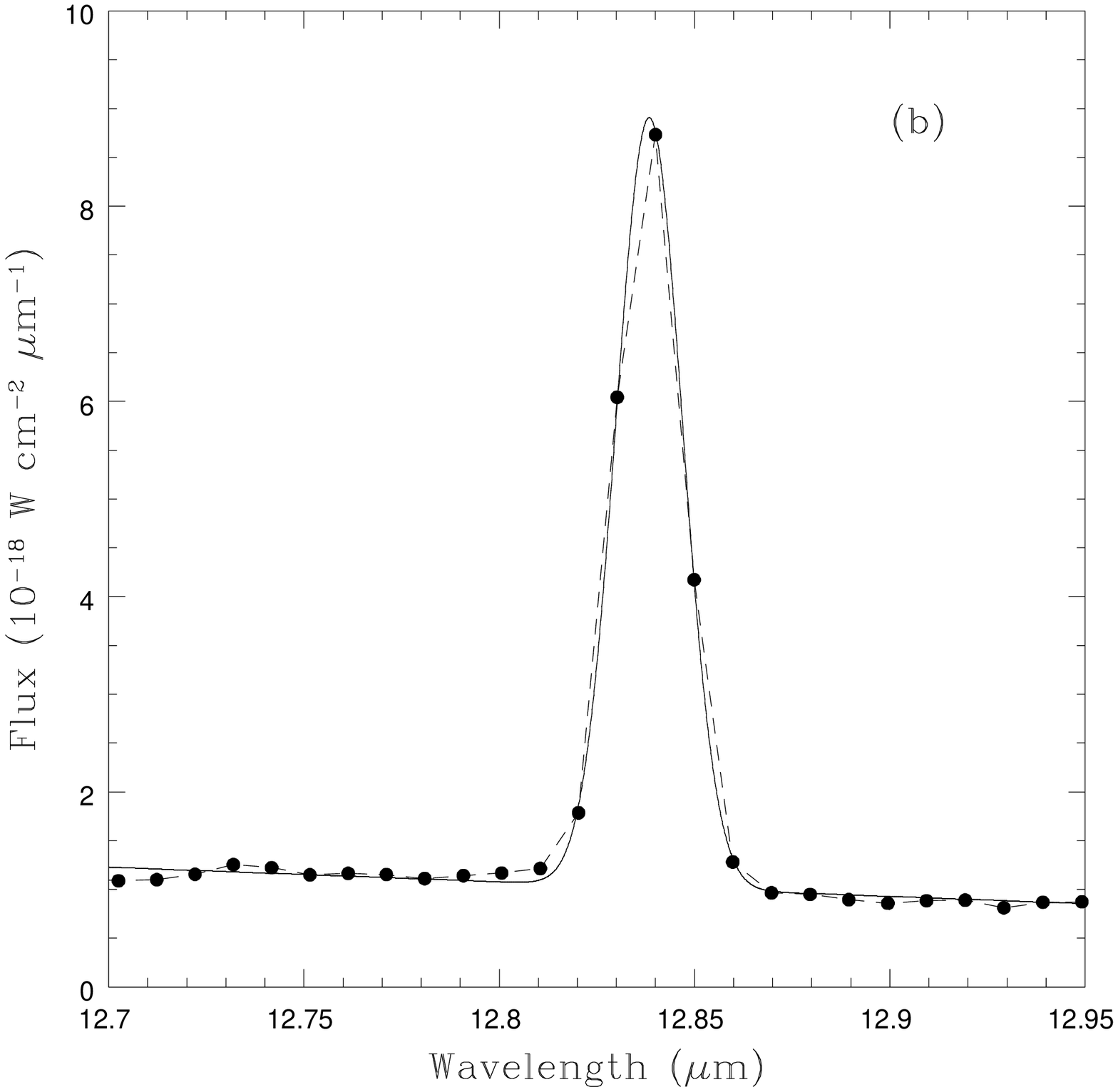}} \\
       \resizebox{80mm}{!}{\includegraphics{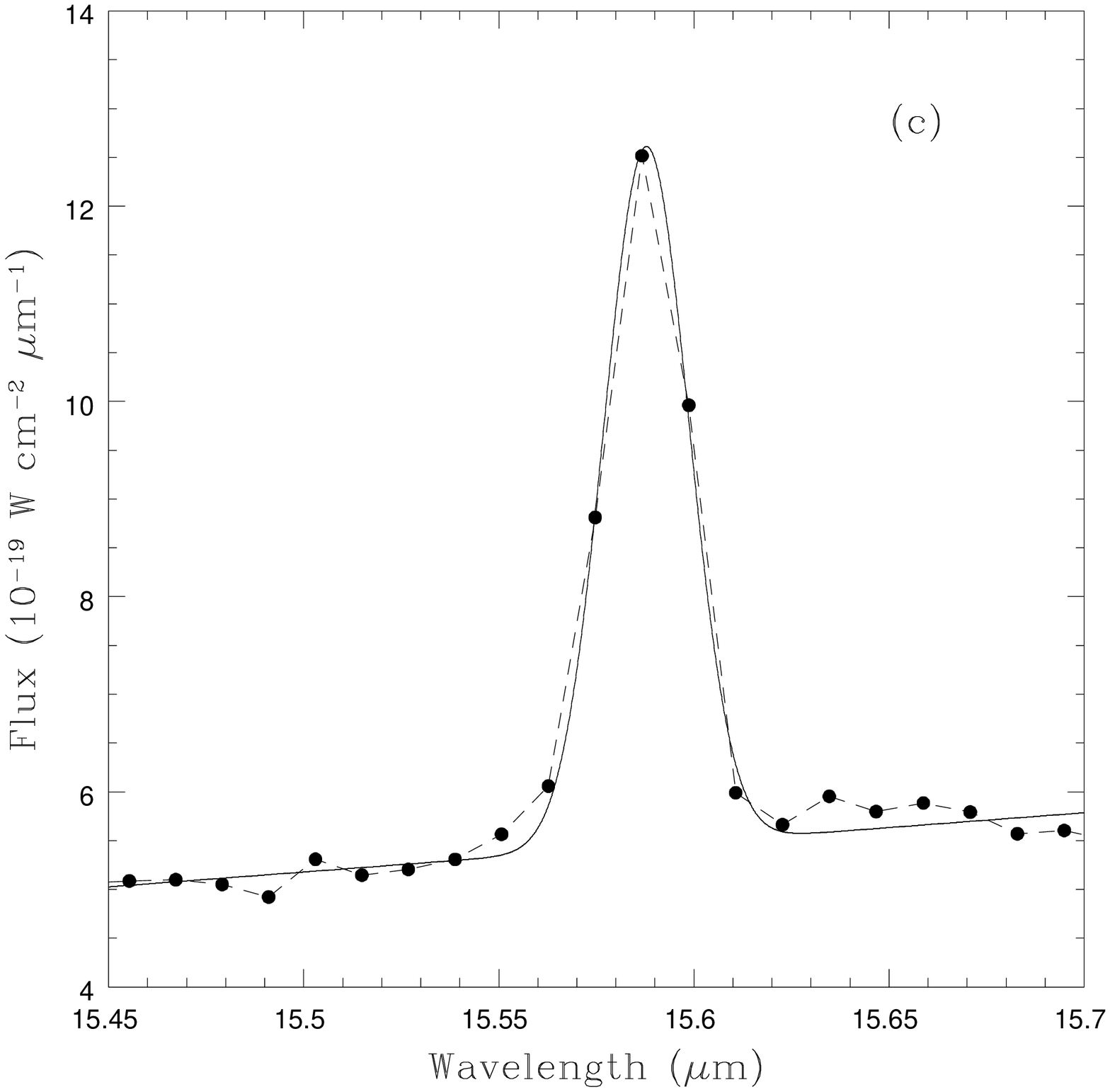}} &
       \resizebox{80mm}{!}{\includegraphics{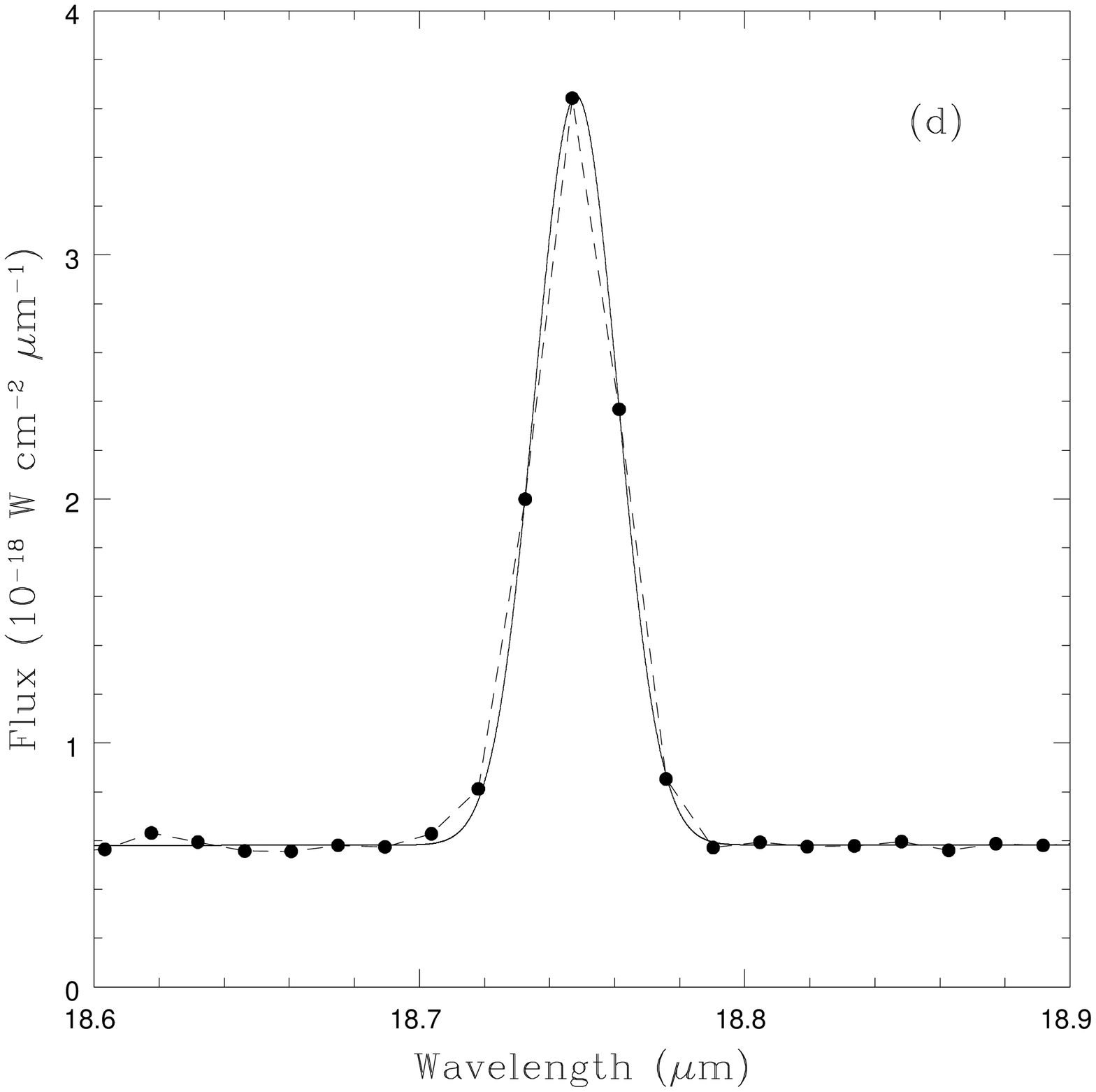}} \\
    \end{tabular}
    \caption{Measurements of the four emission lines in the \ion{H}{2} region
       RK209 (\#9 in Fig.~1): {\bf (a)} [\ion{S}{4}] 10.5~$\mu$m;  
       {\bf (b)} [\ion{Ne}{2}] 12.8~$\mu$m;
       {\bf (c)} [\ion{Ne}{3}] 15.6~$\mu$m; and 
       {\bf (d)} [\ion{S}{3}] 18.7~$\mu$m.
       The data points are the filled circles.
       The fits to the continuum and Gaussian profiles are the solid lines.
       Such measurements provide the set of line fluxes for further analysis.}
    \label{test4}
  \end{center}
\end{figure}

\begin{figure}
\centering
\resizebox{14.0cm}{!}{\includegraphics{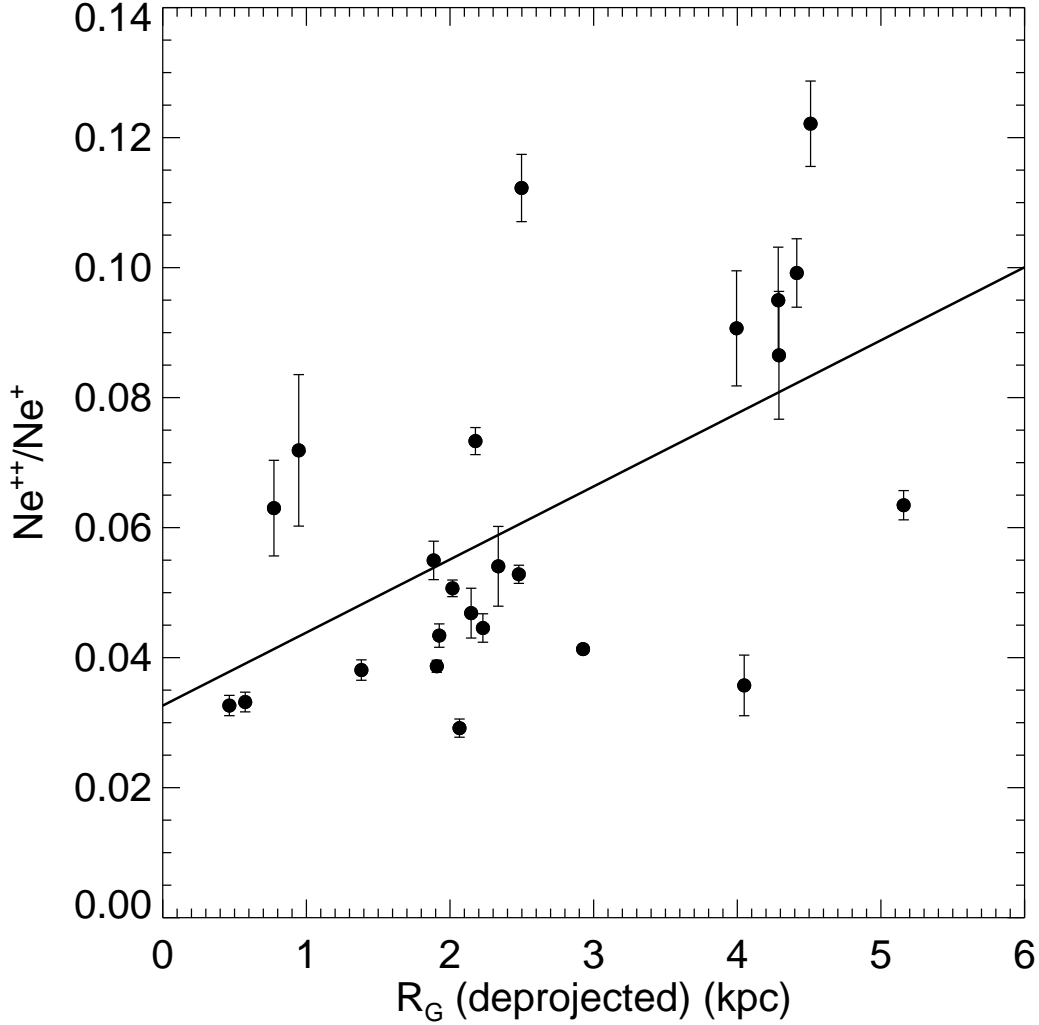} }
\vskip0.1truein
\caption[]{Plot of the ionic abundance ratio 
Ne$^{++}$/Ne$^+$, which is 
derived from the measured line flux ratios for each 
\ion{H}{2} region,  vs.\ R$_G$.
We assume an electron density ($N_e$)
of  100~cm$^{-3}$.
There is extremely little change with $N_e$ over the range
expected for these regions   (see text).
The linear least-squares fit indicates a positive
correlation with R$_G$.
Error bars here and in Figures 4 and 5 are for the propagated measurement 
uncertainties and do not include the systematic uncertainties (see text).}

\end{figure}

\begin{figure}
\centering
\resizebox{14.0cm}{!}{\includegraphics{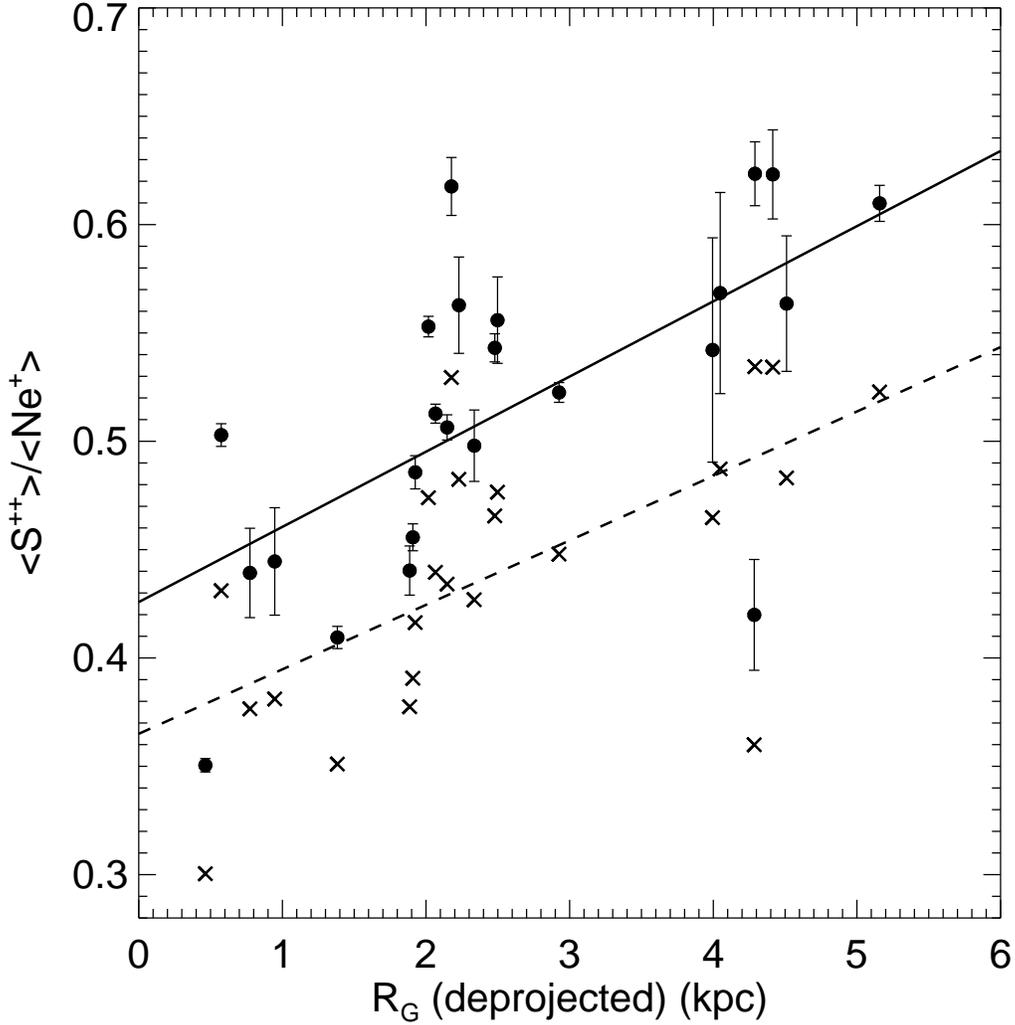} }
\vskip0.1truein
\caption[]{
Plot showing the 
fractional ionic abundance ratio  $<$S$^{++}$$>$/$<$Ne$^+$$>$
vs.\ R$_G$. 
Circles  represent the points and the solid line the
linear least-squares fit for
an assumed $N_e$ of 100~cm$^{-3}$.
For \ion{H}{2} regions
in M83 this is likely the typical
$N_e$ value with an upper limit of roughly 1000~cm$^{-3}$. 
We plot with an X the points and a dashed line the least-squares fit for
an assumed $N_e$~= 1000~cm$^{-3}$,
thereby showing the effect of higher 
$N_e$ on the volume emissivity of the [\ion{S}{3}] line.
The plotted 
$<$S$^{++}$$>$/$<$Ne$^+$$>$ ratio
assumes an Orion Nebula Ne/S 
abundance ratio of 14.3 (Simpson et~al.\ 2004).
Because Ne and S are ``primary" elements,
their production is expected to vary in lockstep
and Ne/S would not be expected to show a radial gradient within a galaxy
(Pagel \& Edmunds 1981).
There is less scatter about the fit than for 
the 
Ne$^{++}$/Ne$^+$~ ratio 
because the 
[\ion{Ne}{3}] 
line is weaker than either
the [\ion{Ne}{2}] 
or [\ion{S}{3}] 
line in these objects.
Error bars are shown for the solid points only to avoid congestion.}

\end{figure}

\begin{figure}
\centering
\resizebox{14.0cm}{!}{\includegraphics{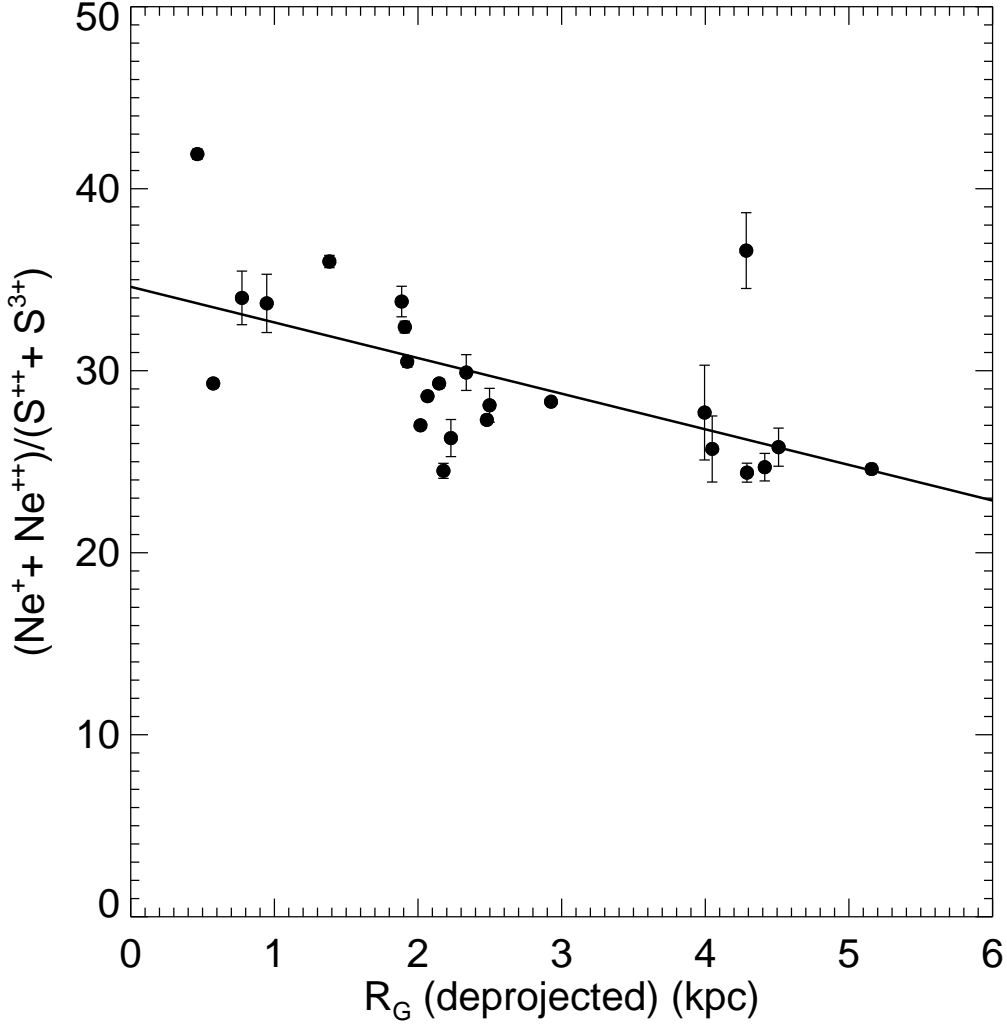} }
\vskip0.2truein
\caption[]{Ne/S, as approximated by
(Ne$^+$ + Ne$^{++}$)/(S$^{++}$ + S$^{3+}$)
(see text) vs.\ R$_G$.
These ratios are all larger than the
Orion Nebula value of 14.3.
There also appears to be a drop in the Ne/S ratio
with increasing R$_G$.
However we do not account for S$^+$,
which must exist in these low ionization M83 \ion{H}{2} region.
Accounting for S$^+$ would lower our estimate of Ne/S.
The expected increasing fraction of
S$^+$ towards the inner galaxy regions would also
lead to a flatter gradient.
Another factor that could lower all the derived Ne/S values, 
as well as flatten the slope, is the higher dust content
(with S, but  not Ne, entering grains)
expected in the inner regions due to higher metallicity.}

\end{figure}

\begin{figure}
  \begin{center}
    \begin{tabular}{cc}
     \resizebox{80mm}{!}{\includegraphics{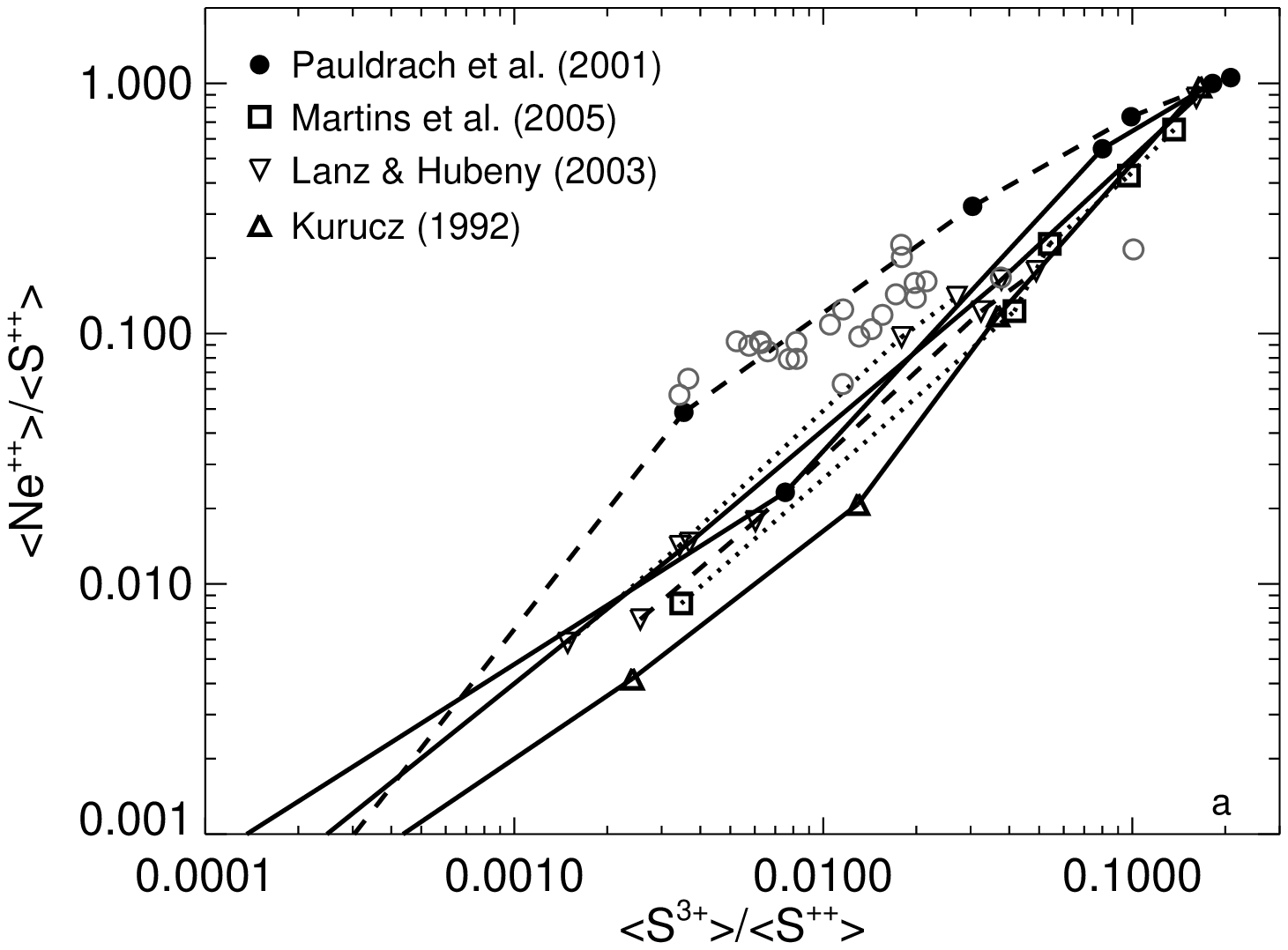}} &
     \resizebox{80mm}{!}{\includegraphics{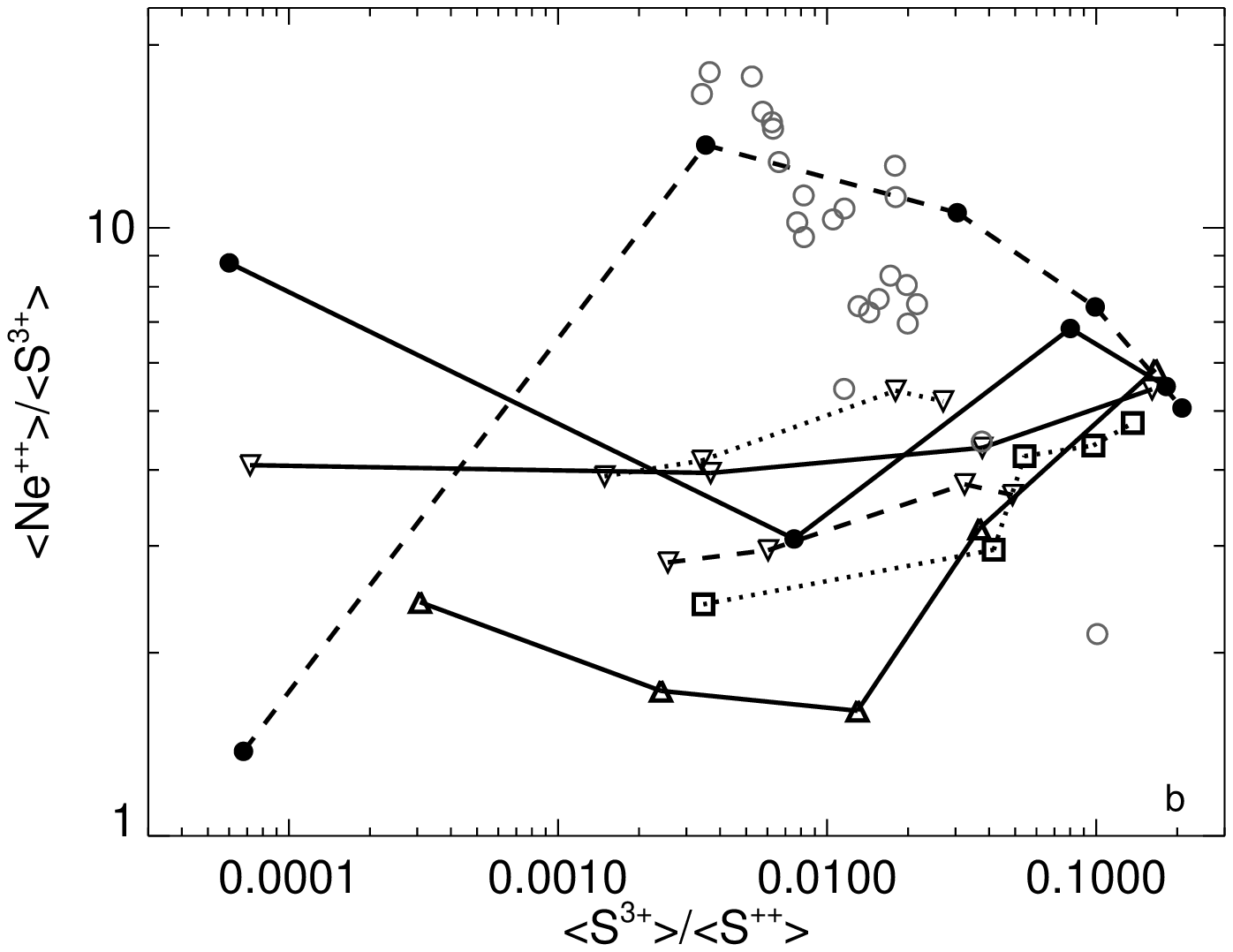}} \\
    \end{tabular}
    \caption[]{{\it {\bf a)}}
Theoretical predictions of the fractional ionization ratios
$<$Ne$^{++}$$>$/$<$S$^{++}$$>$ 
vs.\ 
$<$S$^{3+}$$>$/$<$S$^{++}$$>$,
computed using our photoionization code NEBULA.
The lines connect the results of nebular models calculated with the 
ionizing SEDs predicted from various stellar atmosphere models as labeled,
changing \underbar{no other parameter except the SED}.
For the \ion{H}{2} region models calculated with 
Pauldrach~et~al.\ (2001) atmospheres, 
the solid line connects models with dwarf atmospheres and the dashed line 
connects models with supergiant atmospheres.
For the \ion{H}{2} region models calculated with Lanz \& Hubeny atmospheres,
the solid line connects models with atmospheres with log~$g= 4.0$
and the dotted and dashed lines connect models with atmospheres with
log~$g= 3.0$ to 3.5, and with Lyman continuum luminosities of $10^{49}$
and $10^{50}$ photons s$^{-1}$, respectively.
To compare our data with the models, we need to divide the
observed 
Ne$^{++}$/S$^{++}$
and
Ne$^{++}$/S$^{3+}$
ratios by 
an assumed Ne/S abundance ratio.
We use the Orion Nebula Ne/S~= 14.3.
The open circles 
(adjusted by the assumed Ne/S)
are derived from our observed line fluxes
using $N_e$ of 100~cm$^{-3}$.\break
{\it {\bf b)}} 
The same as panel  a) except
the ordinate is $<$Ne$^{++}$$>$/$<$S$^{3+}$$>$.
Both panels dramatically
illustrate the sensitivity of the H~{\sc ii}
region model predictions of these
ionic abundance ratios to the
ionizing SED that is input to nebular
plasma simulations. 
The M83 data, for the most part, appear to
lie closest to the Pauldrach~et~al.\ 
supergiant loci.}
    \label{FIG6}
  \end{center}

\end{figure}

\begin{figure}
  \begin{center}
    \begin{tabular}{cc}
     \resizebox{80mm}{!}{\includegraphics{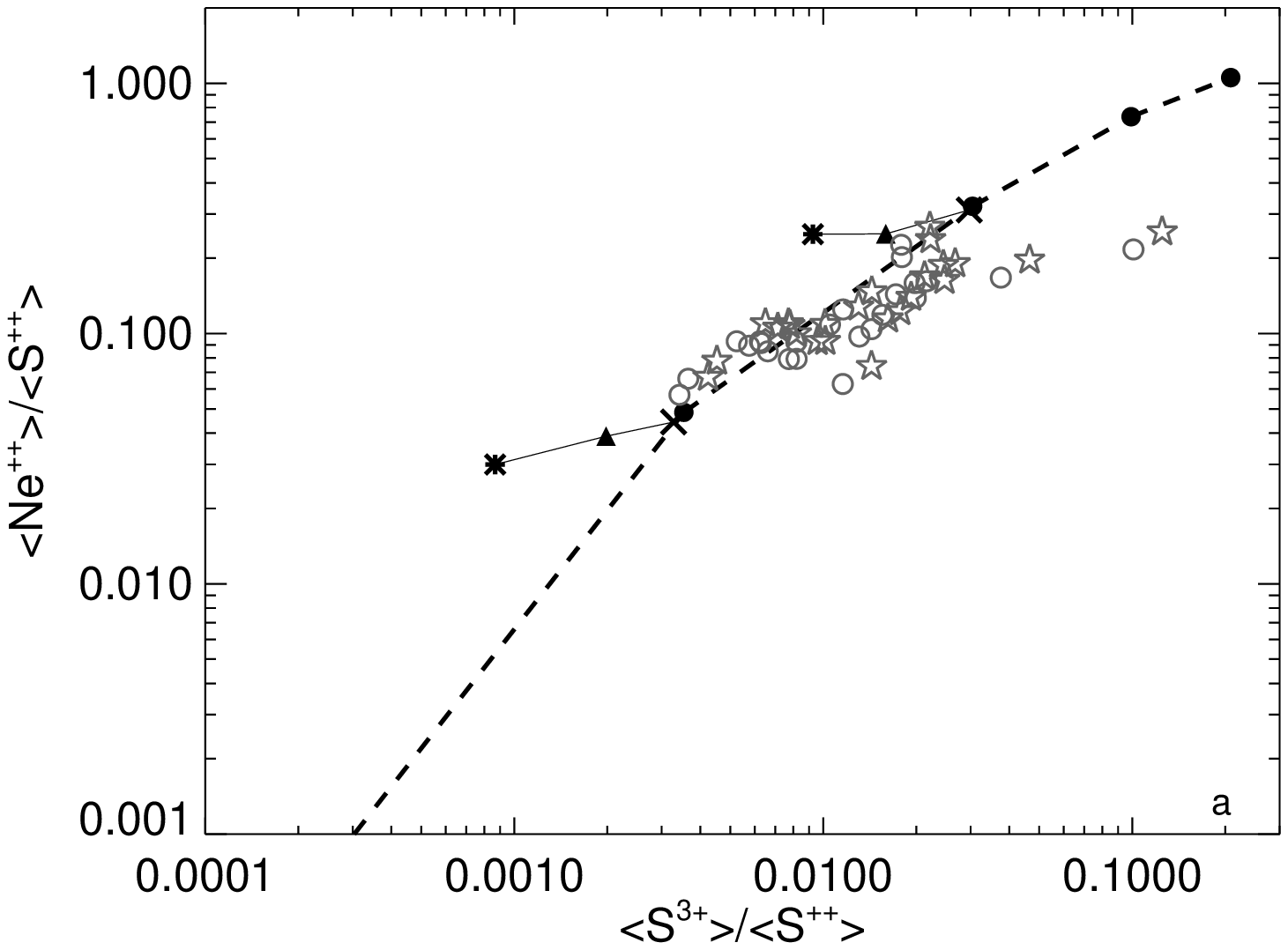}} &
     \resizebox{80mm}{!}{\includegraphics{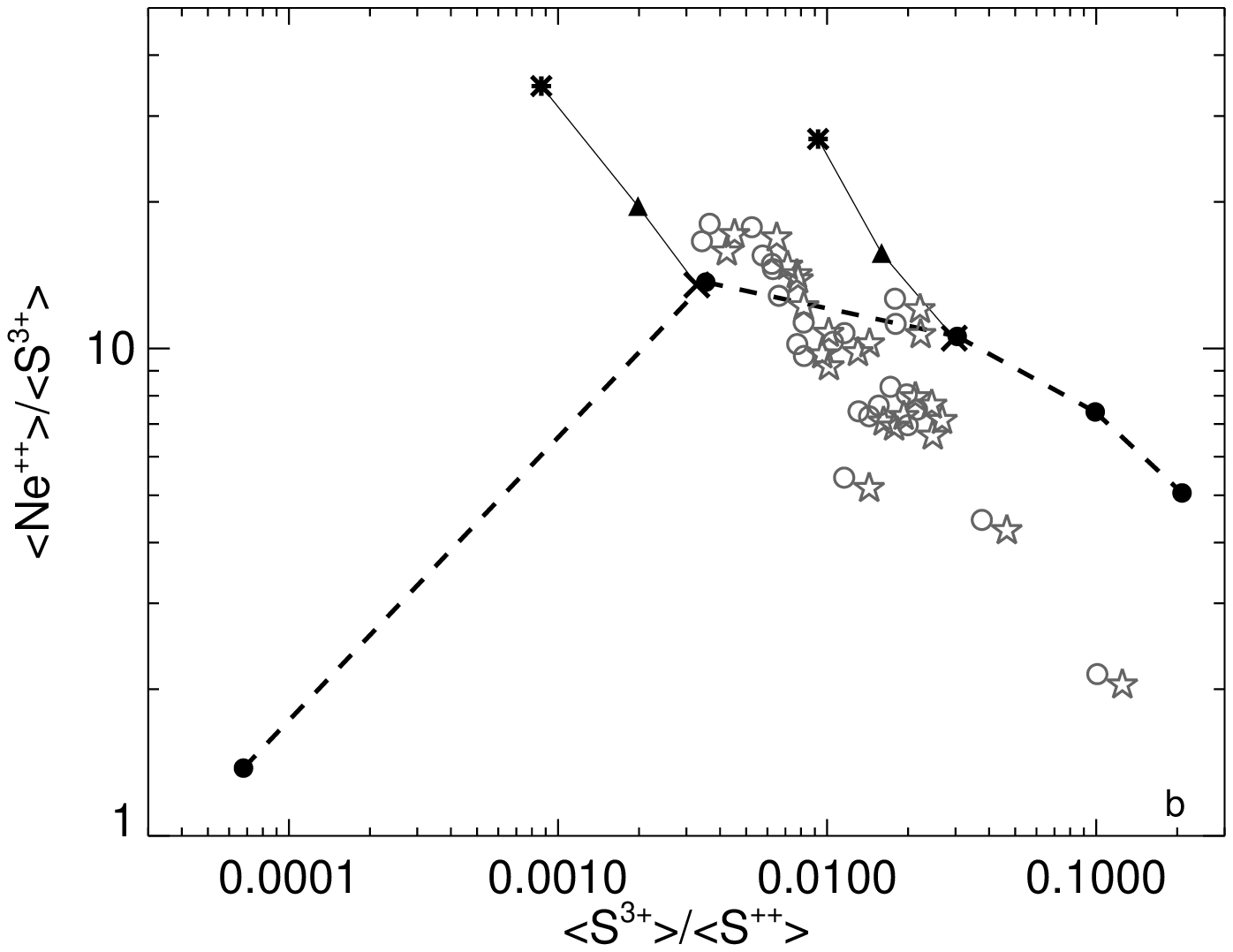}} \\
    \end{tabular}
\caption[]{{\it {\bf a)}}
This is similar to Figure~6a.
We again show the locus that uses the 
Pauldrach~et~al.\ (2001) supergiant atmospheres.
Here we display the results of making some changes
to the {\it nebular parameters} for the \teff\ $=$ 35000 and 40000~K
stellar atmospheres.
The points with an {\bf *} are for a model with a central cavity
of radius 0.5~pc (see text);
those with a triangle have a density of 100 instead
of 1000~cm$^{-3}$;
and those with an {\bf X} have 
a larger number of 
Lyman continuum photons~s$^{-1}$
($N_{Lyc}$)~= 10$^{50}$ instead of 10$^{49}$.
Points corresponding to the same \teff\ are connected by the solid lines.
Again, the open circles are derived from our observed line fluxes
using $N_e$ of 100~cm$^{-3}$.
The open stars show the results of using 
$N_e$ $=$ 1000~cm$^{-3}$~--
a slight shift of the points to the upper right.\break
{\bf b)} 
This is similar to Figure~6b with the same modifications 
as described for panel a).
As shown by the open stars, an $N_e$~= 1000~cm$^{-3}$ 
(compared with $N_e$~= 100~cm$^{-3}$) 
shifts the points slightly to the lower  right.}
    \label{FIG7}
  \end{center}

\end{figure}

  %% 10/31/06 RR - round DEC to nearest arcsec
\begin{deluxetable}{ccccc}
% \begin{deluxetable}{ccccccccccccc}
% \rotate
\tabletypesize{\scriptsize}
\tablecaption{H~{\sc ii} Regions Observed in M83 \label{tbl-1}}
\tablewidth{0pt}
\tablehead{
\colhead{Order} & \colhead{H~{\sc ii} Region} 
& \colhead{~~~~~~~~~~~~~~~~~RA ~~~~~~J2000} & \colhead{DEC} &
\colhead{Aperture Grid}
}
\startdata
% Object          RA            DEC      1907(m,ns) 1 sigma(ns) 1909(m,ns) 3 sigma(ns) 1910(g,ns) FWHM(m) 1907/1909  Ne 12C/13C
% Order          HII            RA                 DEC             Aperture Grid
1  & RK275 & 13 36 40.3 & -29 51 21 & 1x2 \\

2  & RK268 & 13 36 42.5 & -29 52 09 & 1x2 \\

3  & RK266 & 13 36 43.4 & -29 52 22 & 2x2 \\
 
4  & RK230 & 13 36 50.7 & -29 52 02 & 1x3 \\

5  & deV10 & 13 36 52.3 & -29 53 00 & 2x3 \\

6  & RK213 & 13 36 52.6 & -29 51 46 & 1x3 \\

7  & deV13 & 13 36 52.7 & -29 52 46 & 2x3 \\
 
8  & RK211 & 13 36 52.9 & -29 51 11 & 2x4 \\

9  & RK209 & 13 36 53.2 & -29 51 31 & 2x4 \\

10 & RK201 & 13 36 53.9 & -29 48 54 & 2x4 \\

11 & RK198 & 13 36 54.3 & -29 50 47 & 2x4 \\

12 & deV22 & 13 36 54.7 & -29 53 05 & 2x4 \\

13 & RK172 & 13 36 57.1 & -29 51 55 & 2x2 \\

14 & RK154 & 13 36 58.7 & -29 48 06 & 1x2 \\

15 & deV28 & 13 36 59.0 & -29 51 26 & 1x3 \\

16 & deV31 & 13 37 00.0 & -29 52 19 & 1x3 \\

17 & RK137 & 13 37 01.4 & -29 51 27 & 2x4 \\

18 & RK135 & 13 37 02.0 & -29 55 31 & 2x2 \\

19 & RK120 & 13 37 03.5 & -29 54 02 & 1x3 \\

20 & RK110 & 13 37 04.7 & -29 50 58 & 2x3 \\

21 & RK86 & 13 37 07.1 & -29 49 36 & 1x3 \\

22 & RK69 & 13 37 08.5 & -29 52 04 & 1x3 \\

23 & deV52+RK70 & 13 37 08.6 & -29 52 11 & 1x4 \\

24 & RK20 & 13 37 16.9 & -29 53 14 & 2x3 \\

\enddata
\end{deluxetable}

%\include{table1.tex}

  %% 10/31/06 RR - correct 1 number ... 1 sigma error for RK135 Ne III to  2.40E-22
\begin{deluxetable}{ccccccc}
%\rotate
\tabletypesize{\scriptsize}
\tablecaption{M83 Line Measurements\label{tbl-2}}
\tablewidth{0pt}
\tablehead{
\colhead{Order} & \colhead {Source} & \colhead{Line} & \colhead{Flux}
& \colhead{1$\sigma$ error} & \colhead{FWHM} & \colhead{V$_r$}
\\ 
& & \colhead{$\mu$m} & \colhead{W cm$^{-2}$} & \colhead{W cm$^{-2}$}
& \colhead{km s$^{-1}$} & \colhead{km s$^{-1}$}

%%%%%%%%%%%%%%%%%%%%%%%%%%%%%%%%%%%%%%%%%%%%%
%\colhead{Object} & \colhead{~~RA ~~~J2000} & \colhead{DEC} & \colhead{F(1906.7)} &
%\colhead{1$\sigma$ error} & \colhead{F(1908.7)} & \colhead{3$\sigma$}  
%& \colhead{F(1909.6)} & \colhead{FWHM} & \colhead{1907/1909} &\colhead{$N_e$ cm$^{-3}$}
%& \colhead{$r$ ($^{12}$C/$^{13}$C)}
%%%%%%%%%%%%%%%%%%%%%%%%%%%%%%%%%%%%%%%%%%%%%
}
\startdata

% Order Source  Line  Flux  1sigma error   fwhm   vr
1 &  RK275  & 10.5 & 9.73E-22 & 1.19E-22 &  360 & 454\\
  &         & 12.8 & 2.25E-20 & 1.96E-22 &  510 & 627\\
  &         & 15.6 & 3.27E-21 & 1.12E-22 &  493 & 680\\
  &         & 18.7 & 1.44E-20 & 1.56E-22 &  501 & 610\\

2 &  RK268  & 10.5 & 1.15E-21 & 2.16E-22 &  809 & 613\\
  &         & 12.8 & 4.06E-21 & 1.60E-22 &  504 & 651\\
  &         & 15.6 & 1.14E-21 & 4.15E-23 &  456 & 693\\
  &         & 18.7 & 2.41E-21 & 9.71E-23 &  465 & 606\\

3 &  RK266  & 10.5 & 1.62E-21 & 4.20E-22 &  711 & 650\\
  &         & 12.8 & 2.64E-20 & 3.12E-22 &  504 & 657\\
  &         & 15.6 & 5.22E-21 & 5.90E-22 &  502 & 690\\
  &         & 18.7 & 1.73E-20 & 3.65E-22 &  496 & 631\\

4 &  RK230  & 10.5 & 8.14E-22 & 1.22E-22 &  385 & 417\\
  &         & 12.8 & 1.65E-20 & 2.47E-22 &  526 & 638\\
  &         & 15.6 & 4.23E-21 & 1.84E-22 &  513 & 694\\
  &         & 18.7 & 9.61E-21 & 3.23E-21 &  505 & 629\\

5 &  deV10  & 10.5 & 2.13E-21 & 1.22E-22 &  290 & 467\\
  &         & 12.8 & 8.41E-20 & 6.42E-22 &  483 & 674\\
  &         & 15.6 & 1.02E-20 & 1.15E-21 &  517 & 718\\
  &         & 18.7 & 4.30E-20 & 1.43E-21 &  498 & 653\\

6 &  RK213  & 10.5 & 7.40E-22 & 1.23E-22 &  392 & 508\\
  &         & 12.8 & 8.51E-20 & 2.78E-22 &  476 & 654\\
  &         & 15.6 & 5.68E-21 & 2.72E-22 &  479 & 697\\
  &         & 18.7 & 4.59E-20 & 3.73E-22 &  487 & 635\\

7 &  deV13  & 10.5 & 2.01E-21 & 2.66E-22 &  293 & 461\\
  &         & 12.8 & 9.80E-20 & 7.14E-22 &  472 & 672\\
  &         & 15.6 & 1.05E-20 & 8.54E-22 &  502 & 706\\
  &         & 18.7 & 5.21E-20 & 4.77E-22 &  491 & 648\\

8 &  RK211  & 10.5 & 2.30E-21 & 2.25E-22 &  326 & 434\\
  &         & 12.8 & 1.01E-19 & 1.15E-21 &  499 & 624\\
  &         & 15.6 & 1.03E-20 & 4.90E-22 &  529 & 665\\
  &         & 18.7 & 5.95E-20 & 2.32E-21 &  495 & 605\\

9 &  RK209  & 10.5 & 2.90E-21 & 3.25E-22 &  404 & 403\\
  &         & 12.8 & 1.68E-19 & 1.04E-21 &  467 & 580\\
  &         & 15.6 & 1.95E-20 & 4.78E-22 &  491 & 631\\
  &         & 18.7 & 9.77E-20 & 5.97E-22 &  478 & 563\\

10&  RK201  & 10.5 & 2.79E-21 & 4.19E-22 &  495 & 434\\
  &         & 12.8 & 2.76E-20 & 4.30E-22 &  484 & 566\\
  &         & 15.6 & 5.73E-21 & 5.53E-22 &  906 & 587\\
  &         & 18.7 & 1.58E-20 & 1.53E-21 &  486 & 567\\

11&  RK198  & 10.5 & 3.45E-21 & 3.99E-22 &  548 & 362\\
  &         & 12.8 & 7.25E-20 & 1.02E-21 &  479 & 549\\
  &         & 15.6 & 1.22E-20 & 3.01E-22 &  516 & 587\\
  &         & 18.7 & 4.71E-20 & 8.02E-22 &  487 & 528\\

12&  deV22  & 10.5 & 2.99E-21 & 4.76E-22 &  716 & 280\\
  &         & 12.8 & 2.00E-19 & 1.91E-21 &  490 & 512\\
  &         & 15.6 & 1.77E-20 & 3.91E-22 &  469 & 655\\
  &         & 18.7 & 9.60E-20 & 9.62E-22 &  491 & 595\\

13&  RK172  & 10.5 & 4.67E-22 & 1.20E-22 &  343 & 539\\
  &         & 12.8 & 1.03E-20 & 3.00E-22 &  514 & 605\\
  &         & 15.6 & 1.57E-21 & 2.57E-22 &  431 & 658\\
  &         & 18.7 & 4.59E-21 & 2.21E-22 &  437 & 601\\

14&  RK154  & 10.5 & 6.78E-22 & 9.92E-23 &  300 & 385\\
  &         & 12.8 & 1.11E-20 & 1.50E-22 &  502 & 551\\
  &         & 15.6 & 2.52E-21 & 1.30E-22 &  550 & 570\\
  &         & 18.7 & 7.29E-21 & 2.26E-22 &  506 & 530\\

15&  deV28  & 10.5 & 5.21E-22 & 1.51E-22 &  497 & 407\\
  &         & 12.8 & 1.39E-20 & 2.81E-22 &  488 & 563\\
  &         & 15.6 & 2.01E-21 & 2.31E-22 &  809 & 593\\
  &         & 18.7 & 6.44E-21 & 2.82E-22 &  470 & 560\\

16&  deV31  & 10.5 & 1.04E-21 & 7.98E-23 &  294 & 463\\
  &         & 12.8 & 1.14E-19 & 6.59E-22 &  496 & 621\\
  &         & 15.6 & 8.49E-21 & 4.04E-22 &  542 & 661\\
  &         & 18.7 & 4.19E-20 & 2.84E-22 &  505 & 602\\

17&  RK137  & 10.5 & 1.56E-21 & 3.67E-22 &  415 & 342\\
  &         & 12.8 & 1.71E-19 & 1.09E-21 &  488 & 518\\
  &         & 15.6 & 1.30E-20 & 5.88E-22 &  557 & 570\\
  &         & 18.7 & 9.05E-20 & 7.71E-22 &  493 & 502\\

18&  RK135  & 10.5 & 7.76E-22 & 1.15E-22 &  325 & 531\\
  &         & 12.8 & 2.38E-20 & 1.04E-21 &  495 & 649\\
  &         & 15.6 & 1.95E-21 & 2.40E-22 &  466 & 661\\
  &         & 18.7 & 1.42E-20 & 1.01E-21 &  480 & 629\\

19&  RK120  & 10.5 & 1.99E-21 & 8.66E-23 &  439 & 419\\
  &         & 12.8 & 5.66E-20 & 4.44E-22 &  484 & 604\\
  &         & 15.6 & 6.84E-21 & 1.72E-22 &  474 & 646\\
  &         & 18.7 & 3.23E-20 & 3.00E-22 &  493 & 599\\

20&  RK110  & 10.5 & 1.43E-21 & 3.02E-22 &  458 & 379\\
  &         & 12.8 & 1.13E-19 & 9.78E-22 &  480 & 499\\
  &         & 15.6 & 9.85E-21 & 3.98E-22 &  523 & 526\\
  &         & 18.7 & 4.86E-20 & 4.56E-22 &  476 & 483\\

21&  RK86   & 10.5 & 2.06E-21 & 1.51E-22 &  500 & 337\\
  &         & 12.8 & 1.03E-19 & 6.76E-22 &  498 & 486\\
  &         & 15.6 & 9.74E-21 & 1.44E-22 &  486 & 483\\
  &         & 18.7 & 5.66E-20 & 3.38E-22 &  493 & 451\\

22&  RK69   & 10.5 & 8.41E-22 & 1.89E-22 &  582 & 226\\
  &         & 12.8 & 3.32E-20 & 2.97E-22 &  467 & 534\\
  &         & 15.6 & 4.18E-21 & 2.22E-22 &  516 & 545\\
  &         & 18.7 & 1.54E-20 & 3.84E-22 &  477 & 498\\

23& deV52+RK70 & 10.5 & 9.80E-22 & 1.44E-22 &  352 & 421\\
  &         & 12.8 & 7.08E-20 & 8.77E-22 &  487 & 518\\
  &         & 15.6 & 7.03E-21 & 2.78E-22 &  541 & 555\\
  &         & 18.7 & 3.61E-20 & 3.64E-22 &  496 & 489\\

24&  RK20   & 10.5 & 6.61E-22 & 1.60E-22 &  328 & 371\\
  &         & 12.8 & 1.78E-20 & 4.29E-22 &  485 & 512\\
  &         & 15.6 & 3.87E-21 & 3.19E-22 &  580 & 538\\
  &         & 18.7 & 7.85E-21 & 4.53E-22 &  450 & 499\\
\enddata
\end{deluxetable}

%\include{erik.tb2.tex}

  %% 3/1/07 Matt & Erik cross checked
%% 10/30/06 EKK finish
%% 10/27/06 RR reorder someof columns
\begin{deluxetable}{ccccccccccc}	% 11 columns
\rotate
\tabletypesize{\scriptsize}
\tablecaption{Derived Parameters for the H~{\sc ii} Regions in M83 \label{tbl-1}}
\tablewidth{0pt}
\tablehead{
%%%%%%%%%%%%%%%%%%%%%%%%%%%%%%%%%%%%%%%%%
%%%   First row of column heads   %%%%%%%
%%%%%%%%%%%%%%%%%%%%%%%%%%%%%%%%%%%%%%%%%
% Column 1
  \colhead{Order} 
% Column 2
& \colhead{Source} 
% Column 3
& \colhead{R$_G$} 
% Column 4
& \colhead{{\underbar{Ne$^+$}}}
% Column 5
& \colhead{\underbar{Ne$^{++}$}}
% Column 6
& \colhead{\underbar{Ne$^{++}$}}
% Column 7
& \colhead{\underbar{S$^{3+}$}}
% Column 8
& \colhead{\underbar{Ne}}
% Column 9
& \colhead{\underbar{$<$S$^{++}$$>$}}
% Column 10
& \colhead{\underbar{$<$Ne$^{++}$$>$}}
% Column 11
& \colhead{\underbar{$<$Ne$^{++}$$>$}}
%%%%%%%%%%%%%%%%%%%%%%%%%%%%%%%%%%%%%%%%%
%%%   Second row of column heads   %%%%%%
%%%%%%%%%%%%%%%%%%%%%%%%%%%%%%%%%%%%%%%%%
% Column 1
\\ %%% BLANK %%%
% Column 2
& %%% BLANK %%%
% Column 3
& \colhead{kpc}
% Column 4
& \colhead{S$^{++}$} 
% Column 5
& \colhead{S$^{++}$}
% Column 6
& \colhead{Ne$^+$}
% Column 7
& \colhead{S$^{++}$}
% Column 8
& \colhead{S}
% Column 9
& \colhead{$<$Ne$^+$$>$}
% Column 10
& \colhead{$<$S$^{++}$$>$}
% Column 11
& \colhead{$<$S$^{3+}$$>$}  
}
%%%%%%%%%%%%%%%%%%%%%%%%%%%%%%%%
%%%   Column heads are over %%%%
%%%%%%%%%%%%%%%%%%%%%%%%%%%%%%%%

\startdata
%% C1: Order
%% C2: Source
%% C3: R_G
%% C4: Ne+/S++
%% C5: Ne++/S++
%% C6: Ne++/Ne+
%% C7: S3+/S++
%% C8: Ne/S
%% C9: <S++>/<Ne+>
%% C10: <Ne++>/<S++>
%% C11: <Ne++>/<S3+>

%C1%  C2   %  C3  % C4   %  C5   %   C6   %   C7    %  C8  %  C9   %  C10   % C11 %
1  & RK275 & 5.16 & 23.4$\pm$0.3 & 1.49$\pm$0.05  & 0.0635$\pm$0.0023 & 0.0143$\pm$0.0018  & 24.6$\pm$0.3 & 0.610$\pm$0.008 & 0.104$\pm$0.004  & 7.26$\pm$0.92 \\

2  & RK268 & 4.51 & 25.4$\pm$1.4 & 3.10$\pm$0.17  & 0.122$\pm$0.007  & 0.101$\pm$0.019   & 25.8$\pm$1.0 & 0.564$\pm$0.031 & 0.217$\pm$0.012  & 2.15$\pm$0.41 \\

3  & RK266 & 4.29 & 22.9$\pm$0.5 & 1.98$\pm$0.23  & 0.0865$\pm$0.0098 & 0.0199$\pm$0.0052  & 24.4$\pm$0.5 & 0.623$\pm$0.015 & 0.139$\pm$0.016  & 6.96$\pm$1.96 \\
 
4  & RK230 & 2.50 & 25.7$\pm$0.9 & 2.88$\pm$0.16  & 0.112$\pm$0.005  & 0.0180$\pm$0.0028  & 28.1$\pm$0.9 & 0.556$\pm$0.020 & 0.202$\pm$0.011  & 11.2$\pm$1.8 \\

5  & deV10 & 2.34 & 28.7$\pm$1.0 & 1.55$\pm$0.18  & 0.0540$\pm$0.0061 & 0.0105$\pm$0.0007  & 29.9$\pm$1.0 & 0.498$\pm$0.016 & 0.108$\pm$0.013  & 10.3$\pm$1.3 \\

6  & RK213 & 2.07 & 27.9$\pm$0.2 & 0.813$\pm$0.039 & 0.0292$\pm$0.0014 & 0.00343$\pm$0.00057 & 28.6$\pm$0.2 & 0.513$\pm$0.004 & 0.0569$\pm$0.0028 & 16.6$\pm$2.9 \\

7  & deV13 & 2.15 & 28.2$\pm$0.3 & 1.32$\pm$0.11  & 0.0469$\pm$0.0038 & 0.00818$\pm$0.00109 & 29.3$\pm$0.3 & 0.506$\pm$0.006 & 0.0925$\pm$0.0076 & 11.3$\pm$1.8 \\
 
8  & RK211 & 2.23 & 25.4$\pm$1.0 & 1.13$\pm$0.07  & 0.0446$\pm$0.0022 & 0.00820$\pm$0.00086 & 26.3$\pm$1.0 & 0.563$\pm$0.022 & 0.0792$\pm$0.0048 & 9.66$\pm$1.05 \\

9  & RK209 & 2.02 & 25.8$\pm$0.2 & 1.31$\pm$0.03  & 0.0507$\pm$0.0013 & 0.00630$\pm$0.00071 & 27.0$\pm$0.2 & 0.553$\pm$0.005 & 0.0917$\pm$0.0023 & 14.6$\pm$1.7 \\

10 & RK201 & 4.00 & 26.4$\pm$2.5 & 2.39$\pm$0.32  & 0.0907$\pm$0.0089 & 0.0376$\pm$0.0067  & 27.7$\pm$2.6 & 0.542$\pm$0.052 & 0.167$\pm$0.023  & 4.45$\pm$0.79 \\

11 & RK198 & 2.18 & 23.1$\pm$0.5 & 1.70$\pm$0.05  & 0.0733$\pm$0.0021  & 0.0155$\pm$0.0018  & 24.5$\pm$0.4 & 0.618$\pm$0.013 & 0.119$\pm$0.004  & 7.64$\pm$0.90 \\

12 & deV22 & 1.91 & 31.4$\pm$0.4 & 1.21$\pm$0.03  & 0.0387$\pm$0.0009 & 0.00662$\pm$0.00105 & 32.4$\pm$0.3 & 0.456$\pm$0.006 & 0.0849$\pm$0.0020 & 12.8$\pm$2.1 \\

13 & RK172 & 0.95 & 32.1$\pm$1.8 & 2.31$\pm$0.38  & 0.0719$\pm$0.0117 & 0.0216$\pm$0.0057  & 33.7$\pm$1.6 & 0.445$\pm$0.025 & 0.162$\pm$0.027  & 7.49$\pm$2.27 \\

14 & RK154 & 4.41 & 22.9$\pm$0.8 & 2.27$\pm$0.14  & 0.0992$\pm$0.0053 & 0.0198$\pm$0.0030  & 24.7$\pm$0.8 & 0.623$\pm$0.021 & 0.159$\pm$0.009  & 8.05$\pm$1.25 \\

15 & deV28 & 0.77 & 32.5$\pm$1.5 & 2.05$\pm$0.25  & 0.0630$\pm$0.0074 & 0.0172$\pm$0.0050  & 34.0$\pm$1.5 & 0.439$\pm$0.021 & 0.143$\pm$0.018  & 8.35$\pm$2.60 \\

16 & deV31 & 0.46 & 40.8$\pm$0.4 & 1.33$\pm$0.06  & 0.0326$\pm$0.0016 & 0.00525$\pm$0.00041 & 41.9$\pm$0.3 & 0.350$\pm$0.003 & 0.0931$\pm$0.0045 & 17.7$\pm$1.6 \\

17 & RK137 & 0.57 & 28.4$\pm$0.3 & 0.943$\pm$0.043 & 0.0332$\pm$0.0015 & 0.00366$\pm$0.00086 & 29.3$\pm$0.2 & 0.503$\pm$0.005 & 0.0660$\pm$0.0030 & 18.0$\pm$4.3 \\

18 & RK135 & 4.05 & 25.1$\pm$2.1 & 0.898$\pm$0.127 & 0.0357$\pm$0.0047 & 0.0116$\pm$0.0019  & 25.7$\pm$1.8 & 0.568$\pm$0.046 & 0.0629$\pm$0.0089 & 5.43$\pm$1.05 \\

19 & RK120 & 2.48 & 26.3$\pm$0.3 & 1.39$\pm$0.04  & 0.0528$\pm$0.0014 & 0.0131$\pm$0.0006  & 27.3$\pm$0.3 & 0.543$\pm$0.006 & 0.0973$\pm$0.0026 & 7.43$\pm$0.37 \\

20 & RK110 & 1.38 & 34.9$\pm$0.4 & 1.33$\pm$0.06  & 0.0381$\pm$0.0016 & 0.00624$\pm$0.00132 & 36.0$\pm$0.3 & 0.409$\pm$0.005 & 0.0931$\pm$0.0039 & 14.9$\pm$3.2 \\

21 & RK86  & 2.93 & 27.3$\pm$0.2 & 1.13$\pm$0.02  & 0.0413$\pm$0.0007 & 0.00774$\pm$0.00057 & 28.3$\pm$0.2 & 0.523$\pm$0.005 & 0.0791$\pm$0.0013 & 10.2$\pm$0.8 \\

22 & RK69  & 1.89 & 32.5$\pm$0.8 & 1.78$\pm$0.10  & 0.0550$\pm$0.0030 & 0.0116$\pm$0.0026  & 33.8$\pm$0.8 & 0.440$\pm$0.011& 0.125$\pm$0.007  & 10.8$\pm$2.5 \\

23 & deV52+RK70 
     & 1.92 & 29.4$\pm$0.5 & 1.28$\pm$0.05  & 0.0434$\pm$0.0018 & 0.00576$\pm$0.00085 & 30.5$\pm$0.3 & 0.486$\pm$0.008 & 0.0894$\pm$0.0036 & 15.5$\pm$2.4 \\

24 & RK20  & 4.28 & 34.0$\pm$2.1 & 3.23$\pm$0.32  & 0.0950$\pm$0.0082 & 0.0179$\pm$0.0044  & 36.6$\pm$2.1 & 0.420$\pm$0.026 & 0.226$\pm$0.023  & 12.7$\pm$3.2 \\

\enddata
\end{deluxetable}


\begin{thebibliography}{}


\bibitem{afflerbach} Afflerbach A., Churchwell E., Werner M.W., 1997, 
ApJ, 478, 190

\bibitem{asplund2005} Asplund M., Grevesse N., Sauval A.J., 2005, in Cosmic
Abundances as
Records of Stellar Evolution and Nucleosynthesis, 
ed. T.G. Barnes III, F.N. Bash, ASP Conf. Ser., 336, 25

\bibitem{bahcall2005} Bahcall J.N, Basu S., Serenelli A.M., 2005,
ApJ, 631, 1281

\bibitem{bahcall2006} Bahcall J.N., Serenelli A.M., Basu S., 2006,
ApJS, 165, 400

\bibitem{bresolin2002} Bresolin F., Kennicutt R.C.\ Jr., 2002, ApJ, 572, 838

\bibitem{bresolin2005} Bresolin F., Schaerer D., Gonz\'{a}lez Delgado R.M., 
Stasi\'{n}ska G., 2005, A\&A, 441, 981

\bibitem{chiappini2001} Chiappini C., Matteucci F., Romano D., 2001, ApJ, 554, 1044

\bibitem{chiappini2003} Chiappini C., Romano D., Matteucci F., 2003, MNRAS, 339, 63

\bibitem{churchwell} Churchwell E., Smith L.F., Mathis J., Mezger P.G., 
Huchtmeier W., 1978, A\&A, 70, 719

\bibitem{de vaucouleurs} de Vaucouleurs G., Pence W.D., Davoust E., 1983, ApJS, 53, 17

\bibitem{drake} Drake J.J., Testa P., 2005, Nature, 436, 525

\bibitem{dufour} Dufour R.J., Talbot R.J.\ Jr., Jensen E.B., Shields G.A., 
1980, ApJ, 236, 119

\bibitem{giveon} Giveon U., Sternberg A., Lutz D., Feuchtgruber H., 
Pauldrach A.W.A., 2002, ApJ, 566, 880

\bibitem{henry} Henry R.B.C., Worthey G., 1999, PASP, 111, 919

\bibitem{higdon} Higdon S.J.U., et al.,  2004, PASP, 116, 975

\bibitem{hou} Hou J.L., Prantzos N., Boissier S., 2000, A\&A, 362, 921

\bibitem{houck2004} Houck J.R.,  et al., 2004, ApJS, 154, 18

\bibitem{kurucz} Kurucz R.L., 1992, in IAU Symp.\ 149, Stellar Population of 
Galaxies, ed.\ B.\ Barbuy A.\ Renzini, p.\ 225

\bibitem{lanz} Lanz T., Hubeny I., 2003, ApJS, 146, 417

\bibitem{lawrence} Lawrence A., Rowan-Robinson M.,
Ellis R.S., Frenk C.S., Efstathiou  G., Kaiser N., Saunders W.,
Parry I.R., Xiaoyang Xia, Crawford J., 1999, MNRAS, 308, 897

\bibitem{lodders2003} Lodders K., 2003, ApJ, 591, 1220

\bibitem{martin} Martin P., Roy J.-R., 1994, ApJ, 424, 599

\bibitem{martin-hernandez2002a} Mart\'{\i}n-Hern\'{a}ndez N.L., et al., 
2002a, A\&A, 381, 606

\bibitem{martin-hernandez2002b} Mart\'{\i}n-Hern\'{a}ndez N.L., Vermeij R., 
Tielens A.G.G.M., van der Hulst J.M., Peeters E., 2002b, A\&A, 389, 286

\bibitem{martins} Martins F., Schaerer D., Hillier D.J., 2005, A\&A, 436, 1049

\bibitem{mokiem} Mokiem M.R., Mart\'{\i}n-Hern\'{a}ndez N.L., 
Lenorzer A., de Koter A., Tielens A.G.G.M., 2004, A\&A, 419, 319

\bibitem{morisset} Morisset C., Schaerer D., Bouret J.-C., Martins F., 
2004, A\&A, 415, 577

\bibitem{pagel2001}Pagel B.E.J., 2001, PASP, 113, 137

\bibitem{pagel1981} Pagel B.E.J., Edmunds, M.G., 1981, ARA\&A, 19, 77

\bibitem{pauldrach} Pauldrach A.W.A., Hoffmann T.L., Lennon M., 2001, A\&A, 375, 161

\bibitem{rodriguez} Rodr\'{\i}guez M., Rubin R.H., 2005, ApJ, 626, 900

\bibitem{rolleston} Rolleston W.R.J., Smartt S.J., Dufton P.L., 
Ryans R.S.I., 2000, A\&A, 363, 537

\bibitem{rubin1968} Rubin R.H., 1968, ApJ, 154, 391

\bibitem{rubin1985} Rubin R.H., 1985, ApJS, 57, 349

\bibitem{rubin1995} Rubin R.H., Kunze D., Yamamoto T., 1995, 
Astrophy.\ Appl.\ of Powerful New Atomic Databases, A.S.P.\ Conf.\ Ser.\ 78, p.479


\bibitem{rubin2006} Rubin R.H., Simpson J.P., Colgan S.W.J., Dufour R.J., 
Citron R.I.,  Ray K.L., Erickson E.F., Haas M.R., Pauldrach A.W.A., 2006, 
in {\it Galaxy Evolution Across the Hubble Time},
IAU Symposium 235, Eds.~F.\ Combes \& J.\ Palous, (Cambridge U.~Press) (in press)

\bibitem{rubin1994} Rubin R.H., Simpson J.P., Lord S.D., Colgan S.W.J., 
Erickson E.F., Haas M.R., 1994, ApJ, 420, 772

\bibitem{rudolph} Rudolph A.L., Fich M., Bell G.R., Norsen T., 
Simpson J.P., Haas M.R., Erickson E.F., 2006, ApJS, 162, 346

\bibitem{rumstay} Rumstay K.S., Kaufman M., 1983, ApJ, 274, 611

\bibitem{sandford} Sandford S.A.,  Pendleton Y.J.,  \& Allamandola L.J.,
1995, ApJ, 440, 697

\bibitem{seaton} Seaton M.J, 1959, MNRAS, 119, 81

\bibitem{sellmaier} Sellmaier F., Yamamoto T., Pauldrach A.W.A., 
Rubin R.H., 1996, A\&A, 305, L37

\bibitem{shaver} Shaver P.A., McGee R.X., Newton L.M., Danks A.C., 
Pottasch S.R., 1983, MNRAS, 204, 53

\bibitem{shields} Shields G.A., 2002, RMxAC, 12, 189

\bibitem{simpson1995} Simpson J.P., Colgan S.W.J., Rubin R.H., Erickson E.F., 
Haas M.R., 1995, ApJ, 444, 721 

\bibitem{simpson1990} Simpson J.P., Rubin R.H., 1990, ApJ, 354, 165

\bibitem{simpson2004} Simpson J.P., Rubin R.H., Colgan S.W.J., 
Erickson E.F., Haas M.R., 2004, ApJ, 611, 338

\bibitem{smith} Smith L.J., Norris R.P.F., Crowther P.A., 2002, MNRAS, 337, 1309

\bibitem{stasinska} Stasi\'nska G., Schaerer D., 1997, A\&A, 322, 615

\bibitem{sternberg} Sternberg A., Hoffmann T.L., Pauldrach A.W.A., 2003, ApJ, 599, 1333

\bibitem{thum} Thum C., Mezger P.G., Pankonin V., 1980, A\&A, 87, 269 

\bibitem{timmes1995} Timmes F.X., Woosley S.E., Weaver T.A., 1995, ApJS, 98, 617

\end{thebibliography}
\end{document}